\documentclass[aps,pre,twocolumn,superscriptaddress]{revtex4-1}
\pdfoutput=1

\usepackage{palatino}
\usepackage{amsmath}
\usepackage{amssymb}
\usepackage{graphicx}
\usepackage{dcolumn}
\usepackage{boxedminipage}
\usepackage{verbatim}
\usepackage{xcolor}
\usepackage{comment}
\usepackage{subcaption}
\usepackage[colorlinks=true,citecolor=blue,linkcolor=blue]{hyperref}

\graphicspath{{figures/}}

\newcommand{\bfv}[1]{\mbox{\boldmath{$#1$}}}

\newcommand{\x}{\bfv{x}}

\begin{document}

\title{Statistically optimal continuous free energy surfaces from biased simulations and multistate reweighting}

\author{Michael R. Shirts}
\email{michael.shirts@colorado.edu}
\affiliation{Department of Chemical and Biological Engineering, University of Colorado Boulder, Boulder, CO 80305}

\author{Andrew L. Ferguson}
\email{andrewferguson@uchicago.edu}
\affiliation{Pritzker School of Molecular Engineering, University of Chicago, Chicago, IL 60637}

\date{\today}

\begin{abstract}

\noindent Free energies as a function of a selected set of collective variables are commonly computed in molecular simulation and of significant value in understanding and engineering molecular behavior. These free energy surfaces are most commonly estimated using variants of histogramming techniques, but such approaches obscure two important facets of these functions. First, the empirical observations along the collective variable are defined by an ensemble of discrete observations and the coarsening of these observations into a histogram bins incurs unnecessary loss of information. Second, the free energy surface is itself almost always a continuous function, and its representation by a histogram introduces inherent approximations due to the discretization. In this study, we relate the observed discrete observations from biased simulations to the inferred underlying continuous probability distribution over the collective variables and derive histogram-free techniques for estimating this free energy surface. 
We reformulate free energy surface estimation as minimization of a Kullback-Leibler divergence between a continuous trial function and the discrete empirical distribution and show that this is equivalent to likelihood maximization of a trial function given a set of sampled data. We then present a fully Bayesian treatment of this formalism, which enables the incorporation of powerful Bayesian tools such as the inclusion of regularizing priors, uncertainty quantification, and model selection techniques. We demonstrate this new formalism in the analysis of umbrella sampling simulations for the $\chi$ torsion of a valine sidechain in the L99A mutant of T4 lysozyme with benzene bound in the cavity.
\end{abstract}

\maketitle

\section{Introduction}

The free energy as a function of a selected set of collective variable
is an important observable that is ubiquitous in molecular simulation
studies.  This free energy function is frequently called the ``free
energy profile'', ``free energy surface'' or the ``potential of mean
force.'' There can be subtle differences between these quantities in
certain situations, which we briefly discuss later in this article. In
this article, we will use the terms ``free energy surface'' and ``free
energy surfaces,'' and the abbreviation ``FES'' for both the singular
and the plural, in order to emphasize that the theory holds in more
than a single dimension. However, we will use the term ``free energy
profile'' interchangeably with ``free energy surface'' when the
collective variable has only a single dimension.

The calculation of FES parameterized by a small number of
collective variables is largely motivated by the ``curse of
dimensionality''. Molecular systems are intrinsically exceedingly
high-dimensional (with numbers of degrees of freedom in the tens or
hundreds of thousands), which makes study of the system properties in
the full configuration space of limited use in understanding and
controlling molecular behaviors. Instead, system microstates are
frequently projected into a handful of collective variables motivated
by the physics of the problem at hand, and FES are then constructed
over this reduced dimensional space to further analyze.
Applications of free energy profiles include determining the kinetics
of a reaction using the free energy along the reaction
path~\cite{Chandler:JCP:1978,Northrup:P:1982,Schenter:JCP:2003},
understanding the behavior of collective interactions such as
hydrophobicity
~\cite{SanBiagio:EBJ:1998,Sobolewski:JPCB:2007,Makowski:JPCB:2010},
elucidating transport mechanisms through molecular
pores~\cite{Hub:P:2008,Hub:JACS:2010,Allen:BC:2006,Medovoy:B:2016,Sigg:JGP:2014},
and the parameterization of low-dimensional (generalized) Langevin or
Fokker-Planck equations as effective reduced models of the system
dynamics
\cite{Yang:JMB:2007,Hummer:JCP:2003,Kopelevich:JCP:2005,Rzepiela:JCP:2014,Chiavazzo:P:2014}.

There are a number of ways to estimate FES in these collective variables. One could in theory run a simulation and estimate simply calculate the probability of visiting a representative set of the collective variables using histograms, a kernel density approximation, or averaging the mean force. However, free energy barriers in collective variable space exceeding several $k_BT$ in height---where $k_B$ is Boltzmann's constant and $T$ is temperature---are crossed with exponentially small probability in standard (unbiased) simulations, resulting in non-ergodic kinetic trapping and the inability to sample transition states and mechanisms. 

A number of methods have been proposed to overcome this trapping problem. They
typically involve introducing some form of bias of the underlying free
energy landscape to enhance sampling of low probability (high free
energy) regions and accelerate transitions between high probability
(low free energy) metastable states. For example, one can sample rare
values of the collective coordinate by constraining a simulation along
the collective variable. One can then compute the average value of the
force along the collective variable, and properly (though this is
nontrivial) integrating along the collective variable to obtain the
free energy~\cite{Hartmann:EPJST:2011,denOtter:JCTC:2013,Chipot:JCC:1996,Darve:JCP:2008}. The relationship between the mean force and
the FES is why the FES in one dimension is also referred to as the
``potential of mean force''.

However, perhaps the most popular and straightforward way to perform biased sampling is to run an ensemble of $K$ independent simulations, each of which biases the collective variable using a---usually, but not necessarily, harmonic---biasing potential. Each
biasing potential forces the simulation to spend the majority of its
time visiting locations with specific ranges of the collective variables consistent with the biases.  Assuming sampling orthogonal to the collective variables is sufficiently fast, good sampling of the the thermally-relevant domain of the collective variable can be achieved by tiling collective variable space sufficiently densely with biasing potentials such that neighboring biased simulations sample
overlapping configuration spaces. The unbiased FES can then be determined using a range of mathematical approaches
based in importance sampling~\cite{Shirts:AP:2017, Shirts:JCP:2008, Kumar:JCC:1992, Ferguson:JCC:2017}. Provided the collective variables employed are ``good'' in the sense that they adequately separate out the relevant metastable states, this methodology, which goes by the name umbrella sampling~\cite{Torrie:JCP:1977}. Umbrella sampling is a very straightforward and popular approach that works in as many
dimensions as one can adequately cover the space with biasing potentials
with sufficient configurational overlap. Assuming the potential only
depends on the difference in collective variable from the restraint
point, then the unbiased FES can be estimated by \textit{post hoc} analysis of the collective variable at each frame of each biased simulation trajectory without requiring records of the total energies, forces, or any other information from the simulation~\cite{Kumar:JCC:1992}. 

FES are then typically estimated from either biased or unbiased molecular simulation trajectories using a variant of histogramming techniques, most commonly a type of multiple histogram reweighting technique such as the  weighted histogram analysis technique (WHAM)~\cite{Kumar:JCC:1992}.  However, using histograms obscures two important points about FES reconstruction.  First, the
true distribution of observations along the desired collective
variable or variables in the infinite limit is virtually never
actually a histogram but rather a continuous function, so the process of histogramming inherently introduces unnecessary discretization errors.  Second, what we
actually observe when we perform a simulation is neither a histogram, nor a
continuous function, but a discrete set of delta functions, at the observed
values of the collective variables. Approximating the ``true'' FES attained in the limit of infinite sampling of the discrete observations as a histogram inherently entails a loss of information. Although these errors can be and usually are minimized with careful choice of histogram bin width and sufficient sampling, we can resolve these problems with improved approaches to estimate a continuous FES along collective variables directly from the discrete set of empirical observations collected in the simulations that do not introduce the approximation and information loss that histogramming incurs.

We are certainly not the first to observe the disadvantages of histogramming approaches. A number of recent studies have proposed histogram-free methodologies to estimate FES. Westerlund et al.~\cite{Westerlund:JCTC:2018} presented an
  approach that builds FES based on Gaussian mixture models,
  outperforming histogramming, k-nearest neighbors (kNN) and kernel
  density estimators (KDE). Schofield~\cite{Schofield:JPCB:2017} presented an adaptive parameterization scheme for a variety of different possible continuous functions for FES. Lee and co-workers~\cite{Lee:JCTC:2014,Lee:JCTC:2013} presented a variational approach (variational free energy profile, or vFEP) to minimize likelihoods of observations from trial continuous free energy surfaces. Stecher et al.~\cite{Stecher:JCTC:2014} have discussed reconstructing free energy surfaces from umbrella sampling using Gaussian process regression that comes inherently equipped with uncertainty estimates. Schneider et al.~\cite{Schneider:PRL:2017} discuss fitting higher-dimensional FES using artificial neural networks. The umbrella integration method of K\"{a}ster and Thiel~\cite{Kastner:JCP:2005,Kastner:JCP:2009,Kastner:JCP:2012} constructs the FES by numerical integration of a weighted average of the derivative of the free energy with respect to the order parameter. Meng and Roux presented a multivariate linear regression framework to link the biased probability densities of individual umbrella windows to yield a global free energy surface in the desired collective variables, though it uses histograms for some of the intermediate steps~\cite{Meng:JCTC:2015}. Basner and Jarzynski presented an approach to calculate a smoothly varying correction term to a trial continuous potential of mean force~\cite{Basner:JPCB:2008}.

The present work shares particular similarities with the vFEP approach of Lee and co-workers~\cite{Lee:JCTC:2013,Lee:JCTC:2014} and the adaptive parameterization approach of Schofield~\cite{Schofield:JPCB:2017}, but builds upon and goes beyond these works in two main aspects. First, as we detail in our mathematical development, we use the multistate Bennett acceptance ratio (MBAR) approach to furnish the provably minimum variance estimators of the free energy differences required to align independent biased sampling run, and then use these values to compute the maximum likelihood estimate of the unbiased FES. Second, we show how this approach can easily be placed in a fully Bayesian framework that enables transparent incorporation of Bayesian priors, Bayesian uncertainty quantification, and Bayesian model selection.

In this paper, we establish a mathematical framework to relate a discrete observed empirical distribution determined in a set of biased simulations to the unknown and typically continuous ``true'' free energy surface in the collective variables one would expect in the limit of infinite sampling. We present a Bayesian treatment of this formalism to enable the incorporation of regularizing priors, uncertainty quantification, and model selection techniques. We demonstrate our approach in the analysis of umbrella sampling simulations for the $\chi$ torsion of a valine sidechain in lysozyme L99A with benzene bound in the cavity. The focus of the paper is to present analysis methodology, and so we assume that the data collected from biased simulations is sufficient to provide robust estimates of the FES using reasonable methods. As such, it is our goal to calculate the best estimate of the FES given a set of sampled data from biased simulations, where appropriate definitions of ``best'' are explored within this paper.

Although we do not do so here, we observe that it is possible to use current best estimates of the FES to adaptively direct additional rounds of sampling, thereby iteratively improving and refine the FES. Such adaptive methods include
metadynamics~\cite{Laio:P:2002,Huber:JCMD:1994,Barducci:PRL:2008}, adiabatic free energy dynamics~\cite{Rosso:JCP:2002}, temperature accelerated dynamics~\cite{Sorensen:JCP:2000}, temperature accelerated molecular dynamics~\cite{Maragliano:CPL:2006} / driven adiabatic free energy dynamics~\cite{Abrams:JPCB:2008}, 
adaptive biasing force approaches~\cite{Darve:JCP:2008}, 
variationally enhanced sampling~\cite{Valsson:PRL:2014},
and conformational flooding~\cite{Grubmuller:PRE:1995}. 
This class of method has significant
advantages, such as optimally directing computational effort towards under-sampled regions of collective variable space and efficiently reducing uncertainties in the FES. However, these methods do also have significant additional challenges, such as under-sampling slow degrees of motion, and the problems of analyzing simulations that are history-dependent and thus only asymptotically approach equilibrium sampling.
For the purposes of this paper we will therefore consider only equilibrium sampling as the way to generate biased sampling trajectories for the purposes of FES estimation. However, the approach we present is extensible to any collective variable biasing enhanced sampling technique that generates equilibrium samples, and is independent of the type of shape of biasing potential, as long as the potential is not time-dependent. One could not use this approach with the time-dependent biases in a convergence phase of metadynamics, as it would create uncontrolled biases in the result.

Importantly, we also note that our approach is also applicable to data generated with temperature, restraint, or  Hamiltonian exchange~\cite{Sugita:JCP:2000,Bergonzo:JCTC:2014,Li:JCC:2014,Dickson:JCC:2016,Kastner:WCMS:2011}, or expanded ensemble~\cite{Fenwick:JCP:2003,Chodera:JCP:2011}. The only requirement on the data is that samples are collected at equilibrium with respect to a time-independent (i.e., stationary) probability distribution, and the biased samples cover the range of interest of the collective variable.

\section{Theory: FES estimation from biased sampled data}

First, we must be precise about what is being calculated when we calculate a free energy surface. There are
two different free energies as a function of collective variable that one could
calculate. Hartmann et al.~referred to them as as free energies of the
``conditional'' and ``constrained'' ensembles, or alternately the
``geometric'' and ``thermodynamic'' free energies. The differences between these 
two definitions involve differential volumes around the surface created by the collective variable constraint.
The ``thermodynamic FES'' is defined as 
\begin{equation}
F(\vec{\xi}) = -\ln \int_{R^n} e^{-u(\vec{x})} \delta(\Phi(\vec{x})-\vec{\xi}) d\vec{x}, \label{eq:thermodynamic}
\end{equation}
where the value of the collective variables corresponding to a particular system configuration $\vec{x}$ is defined by a low-dimensional mapping
$\Phi(\vec{x})=\vec{\xi}$, and the integral is over the $n$-dimensional real configuration space of the system. We express energies in terms of reduced quantities, such that $u(\vec{x})
= (k_B T)^{-1} U(\vec{x})$, and similarly for free energies.  This expression sums up the probability when the constraint on $\vec{x}$ is satisfied.
The ``geometric FES'', in contrast, is defined as:
\begin{equation}
F(\vec{\xi}) = -\ln \int_{\Sigma(\vec{\xi})} e^{-u(\vec{x})} d\Omega \label{eq:geometric}
\end{equation}
Where $\Sigma(\vec{\xi})$ is the surface of constant $\vec{\xi}$, and
$d\Omega$ is the phase space volume of this surface, and thus is the
logarithm of probability density of the surface $\Sigma(\vec{\xi})$.
This second quantity has also been termed the Riemannian effective
potential~\cite{Fakharzadeh:JPCL:2016,Goolsby:b:2019}.  Several papers
have laid out the very subtle differences in these two definitions,~\cite{Hartmann:EPJST:2011,Fakharzadeh:JPCL:2016} with an examination
of the coarea formula being perhaps the clearest way to see the
relationship.~\cite{Hartmann:EPJST:2011} The derivatives of both
quantities can still be related to the mean force along the collective
variable, with proper corrections for changes of variables which are
beyond the scope of this
summary\cite{Hartmann:EPJST:2011,denOtter:JCTC:2013}.

Fortunately, these two free energy surfaces are easily related by
transforming the reduced energy $u(\vec{x}) \rightarrow u(\vec{x}) \pm
\ln |J_{\Phi}(\vec{x})|$, where $J_{\Phi}$ is the Jacobian of function
$\Phi(\vec{x})$ that maps $\vec{x}$ to $\vec{\xi}$, evaluated at
$\vec{x}$.~\cite{Hartmann:EPJST:2011}. The positive sign takes the
thermodynamic energy surface to the geometric one, and the negative in
the reverse direction. A non-rigorous argument for this correction,
with some abuse of notation, is to note that $\int f(\vec{x})
\delta(\Phi(\vec{x})-\vec{\xi}) d\vec{x} = \int f(\vec{x})
|J_{\Phi}|^{-1}\delta(\Phi(\vec{x})-\vec{\xi)}) d\vec{\xi}$, where we switch
from integrating the delta function over a volume elements
$\vec{x}$ to volume elements of $\vec{\xi}$ because of the presence
of the function $\Phi$ in the $\delta$ function.

The choice of which free energy surface to use is not always clear.
The ``geometric'' quantity may be more useful for determining
transition barriers and it is invariant to the choice if functional
form in the constraint~\cite{Hartmann:EPJST:2011}, but the proper
choice is beyond the scope of this article. We simply note that once
one decides which quantity to calculate, one can replace $u(\vec{x})$
with a reduced potential with the desired Jacobian correction, and all
the steps we present in this paper follow in either case.
For more details on the effects of choosing coordinate systems and
restraint functional forms, we recommend references~\citenum{Hartmann:EPJST:2011, denOtter:JCTC:2013,Chipot:JCC:1996} and \citenum{Fakharzadeh:JPCL:2016}.

Now we have defined what we wish to calculate, we focus on how to actually estimate this free energy surface from data sampled in a simulation. For clarity of exposition, in the present work we will assume the usual case that the biased simulation data are collected at a single temperature and this temperature is the one at which we wish to estimate the unbiased FES. However, the approach we outline here can be generalized to work with simulations in which the biased simulations are carried out at various temperatures~\cite{Sugita:CPL:1999,Hansmann:CPL:1997,Ferguson:JCC:2017,Chodera:JCP:2011} or Hamiltonians~\cite{Fukunishi:JCP:2002,Kwak:PRL:2005}, performed with multiple simulations of each biasing function that are each carried out with different temperatures or modified Hamiltonians, or even performed without biasing potentials, and we lay out some preliminary equations for these approaches either in the text itself or in the Appendix Section~\ref{sec:other_biasing}.

Consider $K$ umbrella sampling simulations with different biasing potentials tiling a collective variable space and enforcing good sampling of all thermally-relevant system configurations with desired values of the collective variable. Typically, the collective variable is 1--3 dimensional, but the formalism holds for arbitrary dimensionality provided the space can be sufficiently densely sampled and sufficient overlaps achieved between neighboring biased distributions.

The reduced potentials $u_{B,k}$ of these states are written in terms of the
original potential $u(\vec{x})$ as:
\begin{equation}
u_{B,k}(\vec{x}) = u(\vec{x}) + b_k(\Phi(\vec{x})-\vec{\xi}_{0,k}) \label{eq:biased_u}
\end{equation}
where the subscript $k$ indexes the biased simulation, the subscript
$B$ reminds us that the potential is biased, $b_k(\vec{\xi})$ is a
user-defined biasing potential as a function of the collective
variables $\vec{\xi}$ in which the umbrella sampling was performed, and 
the restraint point of the biasing potential in the collective variables is defined by $\vec{\xi}_{0,k}$.  
Most commonly, a harmonic potential is used, though the theory
presented here supports \textit{any} functional form of the bias
function of the collective variables.
The biasing potentials are then chosen so that the set of all simulations with biasing potentials give roughly
equal sampling across the relevant range of $\vec{\xi}$ and neighboring biased simulations share overlap in configurational space.

We note two features of our description of umbrella sampling that are germane to our subsequent mathematical developments. First, we do not use the term ``windows'' as is
frequently done when discussing umbrella sampling, as this word possesses significant ambiguity.  ``Window'' could refer to
either a specific interval of values of the collective variable
$\vec{\xi}$, or it could refer one of the $k$ simulations run with
biasing potential $b_k$. These two concepts are related in that
simulations with a biasing potential generally sample values in a
relatively restricted volume around $\vec{\xi}_{0,k}$, but they are certainly not the same thing.
A biased simulation can, in principle, yield any value of $\vec{\xi}$ (although values far from any of the bias minima are highly unlikely) so the simulation results are not strictly within any finite ``window'' of
$\vec{\xi}$ if run for long enough. 

Second, we do not make the problematic assumption that the free energy
of biasing a particular simulation is equal to the value of the FES at
the restraint point $\vec{\xi}_{0,k}$ of the $k$th biasing
potential. This approximation is often called the ``stiff spring''
approximation~\cite{Park:JCP:2003}, as it assumes the collective variable sampling remains
very close to the equilibrium position $\vec{\xi}_{0,k}$ of the bias.
But the value of the free energy of biasing is a weighted average over
all configurations visited by the biasing potential,
and so this approximation deteriorates with increasingly weak biasing potentials. 
Because one has to include biasing potentials of finite width
to sufficiently sample the entire volume of
$\vec{\xi}$ of interest, there is always a tradeoff between the strength and number of biasing potentials used: fewer
biasing potentials require weaker biases, and weaker biases result in less accurate approximations to the free energy at $\vec{\xi}_{0,k}$ under the ``stiff spring'' approximation.  An analysis of this approximation (in the
non-equilibrium pulling case) can be found
in~\cite{hummer:P:2010a}, but the approach presented in the present work avoids this particular problem.

We also note that the problem of approximating the FES using free
energy of the biasing potential is exacerbated by histogramming---as
is done in WHAM---which introduces \textit{additional} bias into the
free energy calculation itself through binning of the energies as well
as the free energies.~\cite{Fajer:JCC:2009} Any sort of averaging of the FES in each bin can be
problematic because it tends to artificially lower barriers, which are
frequently some of the most critical features of the FES that we wish
to accurately resolve.

Given data from biased simulations, we seek the statistically optimal estimate of the FES over the collective variables $F(\vec{\xi})$. This distribution contains exactly the same information content and is essentially interchangeable with the
unbiased probability distribution $P(\vec{\xi})$. These two quantities are simply related through the logarithm:
\begin{equation}
P(\vec{\xi}) \propto e^{-\beta F(\vec{\xi})} \label{eqn:logP}
\end{equation}
where the constant of proportionality is the integral over the collective variable $n$-dimensional volume.
We will work with whichever of the pair is most natural for the
discussion at hand. The relationship is one of proportionality because the right hand side is unnormalized.
It can be turned into a proper probability density dividing by the integral over $\vec{\xi}$ of
$e^{-\beta F(\vec{\xi})}$, which will give the correct units of length$^{-d}$,
  where $d$ is the dimension of $\vec{\xi}$. It is typically the case in molecular simulation that we work with relative, rather than absolute, free energies, in which case $F(\vec{\xi})$ is only defined up to an arbitrary additive constant. In this case, our estimate of the unbiased probability distribution $P(\vec{\xi})$ is only defined up to an arbitrary multiplicative constant anyway.

When we perform a simulation, we obtain an observed, \textit{empirical}
probability distribution consisting of a set of samples
$\{\vec{x}_n\}_{n=1}^N$ distributed over
the space of our collective variables $\vec{\xi}$, with probability density in the collective coordinates $\vec{\xi}$:
\begin{equation}
p_E(\vec{\xi}|\{\vec{x}_n\}) = \sum_{n=1}^N W(\vec{x}_n)\delta (\Phi(\vec{x}_n)-\vec{\xi}) \label{eqn:PE}
\end{equation}
Where $W(\vec{x}_n)$ are weights associated with each sample.

$P_E(\vec{\xi})$ is the most precise description of our sampled
probability density that we have after a simulation, because it
only involves non-zero probability where we actually have
measurements and has zero probability at values of $\vec{\xi}$ that
are not observed. If we only perform a single, unbiased simulation 
then $W(\vec{x}_n) = 1/N$ for every sample, where
$N$ is the number of samples, since---in continuous space with
arbitrarily high resolution of system configurations and collective
variable mapping---each observation occurs only once. However, as we
describe in the next section, if we have $K$ biased simulations, we
can incorporate data from all $\sum_{k=1}^K N_k = N$ points gathered
over all of the $K$ states to better estimate
$P_E(\vec{\xi})$~\cite{Shirts:JCP:2008}.

\subsection{MBAR and the empirical FES}

The multistate Bennett acceptance ratio (MBAR) is the statistically optimal approach to estimate the reduced free energies $f_k = \int e^{-u_k(\vec{x})} d\vec{x}$, 
from $\{\vec{x}_1,\vec{x}_2,\ldots,\vec{x}_N\}$ observations at $K$ thermodynamic state points~\cite{Shirts:JCP:2008}.
These $K$ thermodynamic states are defined by the reduced potentials $\{u_1,u_2,\ldots,u_K\}$, and we assume that the $\{\vec{x}_n\}_{n=1}^N$ are distributed according to the Boltzmann distribution corresponding to the the reduced potential of the state they are collected from.  With these assumptions, the MBAR estimate for the reduced free energy differences between these $K$ states is~\cite{Shirts:JCP:2008}:
\begin{equation}
e^{-\hat{f}_i} = \sum_{n=1}^{N} \frac{e^{-u_i(\vec{x}_n)}}{\sum_{k=1}^K N_k \, e^{\hat{f}_k - u_k(\vec{x}_n)}} \label{equation:estimator-of-free-energies}
\end{equation}
where $N_k$ is the number of samples taken from state $K$. This system of equations must be solved self-consistently for the
estimated reduced free energies $\hat{f}_i$. Since the reduced free energies are typically only defined up to an additive constant, we usually choose to pin one of the estimated free energies $\hat{f}_i$ equal to some constant value (usually zero) and the rest follow the determined relative free energy differences. We note that MBAR may be considered a binless estimator of free energy  differences that can be derived from WHAM in the limit of zero-width bins~\cite{Shirts:JCP:2008,Tan:JCP:2012,Bartels:CPL:2000}.

After we have solved for these $\hat{f}_i$, then we can calculate the weight $W_i$ of sample $\vec{x}_n$ in any state $i$ as~\cite{Shirts:JCP:2008,Bartels:CPL:2000}:
\begin{equation}
W_i(\vec{x}_n) = \frac{e^{\hat{f}_i-u_i(\vec{x}_n)}}{\sum_{k=1}^K N_k \, e^{\hat{f}_k - u_k(\vec{x}_n)}}
\label{eq:MBARweight}
\end{equation}
The weight $W_i(\vec{x}_n)$ of sample $\vec{x}_n$ at thermodynamic state point $i$ represents the contribution to the average of an observable $A$ in state $i$ under a reweighting from the \textit{mixture distribution}, consisting of all samples collected from all $K$ state points, to the state $i$~\cite{Shirts:AP:2017}. The probability of each sample in the mixture distribution is $p(\vec{x}_n) = \sum_{i=1}^K \frac{N_i}{N}p_i(\vec{x}_n) = \sum_{i=1}^K \frac{N_i}{N} e^{\hat{f}_i-u_i(\vec{x}_n)}$---in other words, simply the average of all of the individual $p_i$ probability distributions weighted by the number of samples $N_i$ drawn from each of the $K$ states.~\cite{Shirts:AP:2017} 
It can be easily checked from eq.~\ref{eq:MBARweight} that the $W_i(\vec{x}_n)$ are normalized such that~\cite{Shirts:JCP:2008}:
\begin{equation}
\sum_{i=1}^K N_i W_i(\vec{x}_n)=1
\label{eq:normal}
\end{equation}
and also from eq.~\ref{equation:estimator-of-free-energies} and eq.~\ref{eq:MBARweight} that~\cite{Shirts:JCP:2008}:
\begin{equation}
\sum_{n=1}^N W_i(\vec{x}_n)=1
\label{eq:normal2}
\end{equation}
The expectation value of the observable $A$ estimated over all samples at all state points may then be written as:
\begin{equation}
\langle A\rangle_i = \sum_{n=1}^{N} W_{i}(\vec{x}_n) A(\vec{x}_n) \label{eqn:obs}
\end{equation}
as discussed in eqs. 9 and 15 of the original MBAR paper~\cite{Shirts:JCP:2008}. We denote the weight of sample $\vec{x}$ as obtained via MBAR in the
\textit{unbiased} state as $W(\vec{x}_n)$, and in each of the $k = 1 \ldots K$ \textit{biased} states as $W_k(\vec{x}_n)$.

By eq.~\ref{eqn:logP}, the exponential of minus the free energy surface $F_i$ in state $i$ is
proportional to a probability density.  By combining eq.~\ref{eqn:logP} and eq.~\ref{eqn:obs} under the particular choice for the observable $A(\vec{x}_n) = \delta\left(\Phi(\vec{x}_n)-\vec{\xi}\right)$, we have within the MBAR framework that:
\begin{equation}
P(\vec{\xi}) = \langle \delta\left(\Phi(\vec{x}_n)-\vec{\xi}\right) \rangle_i = \sum_{n=1}^N W_i(\vec{x}_n)\delta\left(\Phi(\vec{x}_n)-\vec{\xi}\right) \label{eqn:weightedSum}
\end{equation}
where $\Phi(\vec{x})$ maps from the full coordinate space to the lower
dimensional collective variable space of interest.

Eq.~\ref{eqn:weightedSum} makes clear that the MBAR estimate of the probability density as a function of $\vec{\xi}$ 
is a weighted sum of delta functions at the observed points. (Technically, it's
a distribution, not a function, since it is a sum of delta
functions, which are themselves are distributions, but this formal distinction doesn't affect any of the development in this paper.) It is instructive to compare this to the empirical distribution
function when collecting samples from a single state where $W_i(\vec{x}_n) = 1/N$:
\begin{equation}
P(\vec{\xi}) = \frac{1}{N}\sum_{n=1}^N \delta\left(\Phi(\vec{x}_n)-\vec{\xi}\right)
\end{equation}
from which it can be seen that the empirical distribution $P_E(\vec{\xi}|\{\vec{x}_n\})$ generated using MBAR in eq.~\ref{eqn:PE} is a 
\textit{weighted} empirical distribution function using data from all states.

The representation of the empirical probability distribution function $P_E(\vec{\xi}|\{\vec{x}_n\})$
of delta functions has both advantages and disadvantages. Estimating expectation values of
observables that are a function of $\vec{\xi}$ becomes simply a weighted sum over all observations 
\begin{equation}
\langle A \rangle_i = \int A(\vec{x}) P(\vec{x}) d\vec{x} = \sum_{n=1}^N W_i(\vec{x}_n) A(\vec{x}_n). \label{eqn:expect}
\end{equation}
However, it is very complicated to interpret or
visualize this delta function representation.  Neither can we work with this
empirical representation in logarithmic form $F(\vec{\xi}) = - \ln P(\vec{\xi})$ because the logarithm of a sum of delta functions isn't defined, so only the exponential form has a well-defined
mathematical meaning.  Again, we have implicitly put the $F(\xi)$ in reduced form so that it is a pure number. We will maintain this convention throughout the remainder of this paper. To change into real energy units we simply multiply through by $k_B T$ so that $F_{\mathrm{units}} = (k_B T) F$. 

To reiterate, expectations of quantities of interest can be computed  by eq.~\ref{eqn:expect} without recourse to $F(\vec{\xi})$ directly, but representing $F(\vec{\xi})$ as a continuous function is valuable for interpretation and understanding of the underlying molecular FES. 
If we have a continuous probability density, we can then define $F(\vec{\xi})=-\ln P(\vec{\xi})$ up to an arbitrary normalization constant of with dimensions (length)$^d$ required to make the argument of the logarithm unitless. 
We will use $F(\vec{\xi})$  to refer to the unbiased FES and $F_k(\vec{\xi})$ to the biased free energy FES obtained from each of the $k = 1 \ldots K$ biased states. 

Developing statistically optimal representations of $F(\vec{\xi})$ that can be visualized and exploited to understand and engineer molecular behaviors is the key motivator of the remainder of this article.

\subsection{Representations of $F(\vec{\xi})$ as a continuous function}

In most cases, to visualize either a $P(\vec{\xi})$ or $F(\vec{\xi})$, or to use
them in some other type of mathematical modeling, we need to choose
how to represent them as continuous functions. Additionally, in the
infinite sampling limit for molecular systems, they generally
\textit{should} be continuous functions due to the inherent continuity of the distribution supported by non-pathological choices of $\vec{\xi}$. We now proceed to
describe a number of possible choices for continuous representations of $F(\vec{\xi})$. Most of the mathematical machinery that we develop can, in principle, be deployed in arbitrarily high dimensionalities of $\vec{\xi}$, although the capacity to achieve sufficient sampling will always present an issue. We note at appropriate junctures in the text any special considerations that may arise when generalizing to high-dimensional parameterizations.

\textbf{1. Represent the FES at specific locations $\vec{\xi}_0$ as
  the free energy of imposing each of the biasing restraints centered
  at $\vec{\xi}_0$.} Assuming we have well-localized biasing
potentials, then the free energy difference between the biased
simulation and the unbiased simulation can be estimated as the free
energy to restrain the simulation by each of the biasing functions. As
described above, this method entails significant drawbacks in
overestimating valleys and underestimating peaks, and in a lack of
resolution between umbrella centers. We do not pursue this further.

\textbf{2. Create a histogram out of the empirical distribution.} This was
  the default choice made in the \texttt{pymbar} package's
  \texttt{computePMF} function, which has occasionally been erroneously
  called the ``MBAR estimate of the potential of mean force'' in the literature.  As we have
  shown, the use of MBAR is completely independent of the
  determination of the FES, although it can be \textit{used} in
  various algorithms to estimate the FES.

  We can calculate the expectation of the binning function 
  $I_i(\vec{\xi}_i,\delta,\vec{x}) = 1$ if $\Phi(\vec{x}) > (\vec{\xi}_i-\delta/2)$
  and $\Phi(\vec{x}) < (\vec{\xi}_i+\delta/2)$ and $I_i(\vec{\xi}_i,\delta,\vec{x})
  = 0$ otherwise, where the $\vec{\xi}_i$ are the centers of the
  histogram bins and with some abuse of notation $\delta$ denotes the multidimensional bin widths, which---for clarity of exposition---we select to be equal in all dimensions.  The binning function is used to essentially assign a fractional count to each bin
  according to the value of $W(\vec{x}_n)$ for $\vec{x}_n$ within
  the bin.  The free energy surface with $J$ total indicator
  functions:
\begin{equation}
F(\vec{\xi}) = -\ln \sum_{i=1}^{J} \sum_{n=1}^N W(\vec{x}_n)I_i(\vec{\xi}_i,\delta,\vec{x}_n)
\end{equation}
where the second sum, as discussed above, is over all $N$ samples collected from all biased
  simulations. Since we are calculating a log expectation of a function,
  MBAR gives a straightforward estimate for the error in the
  uncertainties, as outlined in the original MBAR
  paper~\cite{Shirts:JCP:2008}. If the bin widths are chosen adaptively with
  the number of samples, the uncertainty becomes more complicated,
  since a different data set would have a different set of
bin widths. If we wished, we could fit this histogram to a smooth
  function, using a least squares fitting method, choosing the function
  to balance variance and bias. However it is better to
  avoid any histogramming steps altogether due to the inherent and potentially uncontrolled bias that they introduce. 
  This is especially true with
  multidimensional histograms, where the curse of dimensionality causes the number of bins required, and
  thus the number of samples for equal resolution, to scale exponentially with dimensionality.   We do emphasize that with sufficient data and attention to histogram bin size, these errors can be minimized, and thus the majority of the free energy surfaces in the literature obtained by histograms are sufficiently accurate for the purposes of their studies.

  When WHAM is employed to perform the FES estimation~\cite{Kumar:JCC:1992}, the histograms used to compute the free energies
  are the same as the ones used to calculate the FES, which has a
  tendency to average out the FES~\cite{Fajer:JCC:2009}. With MBAR,
  one can choose exactly how wide to make the histograms, since the
  histograms can be of any width that one chooses to best represent
  the underlying data, and are not constrained by the choice of
  separation in $\vec{\xi}$ between biasing functions $b_k(\vec{\xi})$~\cite{Shirts:JCP:2008}.

\textbf{3. Employ a kernel density approximation.} We can replace each delta function
  in the empirical FES with a smooth function with weight centered at
  each sample and scaled by the weight.  The most common choice is an isotropic
  Gaussian kernel $K(\vec{\xi}_i,\delta,\vec{\xi}) = (2\pi \delta^2)^{-\frac{1}{2}}
  e^{-\frac{(\vec{\xi}-\vec{\xi}_i)^2}{2\delta^2}}$, where $\delta$ now plays the role of the kernel bandwidth, but anisotropic Gaussians, ``top hat,''
  and triangle functions are also frequently used. We observe that histogramming can be considered a form of kernel density  estimation using indicator functions, with the center of the mass the preassigned bin center rather than the location of the sample.
The bandwidth
  $\delta$ can be calculated in a number of ways, although the optimal
  choice is frequently not obvious~\cite{Park::1992,Cao:CSDA:1994,Jones:JASA:1996,Sheather:JRSSBM:1991}. However, the maximum likelihood approach with the empirical distribution shrinks $\delta$ to zero, so other approaches must be used. The FES in the kernel density approximation then becomes: 
\begin{equation}
F(\vec{\xi}) = -\ln \sum_{n=1}^N W(\vec{x}_n)K(\Phi(\vec{x}_n),\delta,\vec{\xi})
\end{equation}
though to make this well-defined, one should check that the kernels result in probability being defined for all values of $\vec{\xi}$ of interest.

\textbf{4. Identify a parameterized continuous probability distribution that best represents the empirical distribution.} The fundamental difficulty with this approach is that there is no unambiguous ``best'' continuous distribution that stands independent of any other assumptions beyond those made so far. Specifically, the closest parameter-independent continuous function to a set of $\delta$ functions, for any reasonable definitions of close, are continuous functions that are essentially indistinguishable from the $\delta$ functions themselves. It is necessary, therefore, to instead impose some constraints upon the family  of continuous functions that represent our understanding of the empirical distribution as a discrete finite-data sampling of what should be a smooth and continuous distribution in the limit of infinite samples. This extremely flexible point-of-view allows for a variety of ways to represent the function with minimal bias and which naturally admit Bayesian formulations. The examination of this fourth perspective is our focus for the remainder of the paper. We proceed to present a number of possible ``best'' choices for the representation for this continuous function along with proposed quantitative definitions of ``best''.

\subsection{Kullback-Leibler divergence as a measure of distance}

Before we start examining mathematical forms of the trial FES, we need to decide how we
will evaluate how close a (continuous) trial function
$P_T(\vec{\xi}|\vec{\theta})$ of some arbitrary parameters
$\vec{\theta}$ is to the empirical distribution $P_E(\vec{\xi}|\{\vec{x}_n\})$.  For the purposes of the present mathematical development we will leave the form of $P_T(\vec{\xi}|\vec{\theta})$ abstract, but it can be useful to consider that a number of parameterizations for the trial function are possible, including linear interpolants, cubic splines, or piecewise cubic Hermite interpolating polynomial (PCHIP) interpolations.
For non-pathological continuous representations of $P_T(\vec{\xi}|\vec{\theta})$, the
corresponding FES is simply $F(\vec{\xi}|\vec{\theta}) = - \ln P_T(\vec{\xi}|\vec{\theta})$. 

One logical definition of ``closeness'' is the
Kullback-Leibler (KL) divergence from the empirical distribution in
the state of interest (the one without any biasing distribution) to
our trial distribution $P_T(\vec{\xi}|\vec{\theta})$, over the volume
$\Gamma$ of collective variables. The Kullback-Leibler divergence from $Q$ to $P$, denoted
$D_{\mathrm{KL}}(P||Q)$, can be interpreted as a measure of the
information lost when $Q$ is used to approximate $P$, and is
defined as:
\begin{equation}
D_{\mathrm{KL}}(P||Q) = \int_{\Gamma} P(\x) \ln \frac{P(\x)}{Q(\x)} d\x
\end{equation}
In later usage, we will generally omit the explicit reference to the volume $\Gamma$ over the collective variable space. 
We will develop several different formulations of the KL divergence
that each consist of a weighted sum of the function evaluated at each
sampled point, and the integral of the simulation over all the entire
FES (or sum of several integrals).  We present them here and then later report  the results of numerical tests to demonstrate their performance.

\textbf{C.1. Unbiased state Kullback-Leibler divergence.} The KL divergence from $P_T(\vec{\xi}|\vec{\theta})$ to $P_E(\vec{\xi}|\{\vec{x}_n\})$ is:
\begin{eqnarray}
D_{\mathrm{KL}}(\vec{\theta}) &=& \int P_E(\vec{\xi}|\{\vec{x}_n\}) \ln \frac{P_E(\vec{\xi}|\{\vec{x}_n\})}{P_T(\vec{\xi}|\vec{\theta})} d\vec{\xi} \nonumber \\
               &=& \int \left[P_E(\vec{\xi}|\{\vec{x}_n\}) \ln P_E(\vec{\xi}|\{\vec{x}_n\})\right. \nonumber \\
               & & \left.-P_E(\vec{\xi}|\{\vec{x}_n\}) \ln P_T(\vec{\xi}|\vec{\theta})\right] d\vec{\xi} 
\end{eqnarray}
The first term in the integral is somewhat problematic, in that it has
a factor of $\ln P_E(\vec{\xi}|\{\vec{x}_n\})$, which is not well-defined for delta functions. Even taking Gaussian approximations for the delta functions and allowing them to shrink to zero-width fails to yield a well-defined value since the entire integral $\int P_E(\vec{\xi}) \ln P_E(\vec{\xi})$ is unbounded in the positive direction as the width of the $\delta$
  function goes to zero.
Fortunately, whatever the value may be, it is independent of the
parameters $\vec{\theta}$. Accordingly, we may neglect the first term in our minimization with respect to $\vec{\theta}$ and focus only on minimization of the second term. For the purposes of functional optimization we will---with some abuse of terminology---use
  $D_{\mathrm{KL}}(\vec{\theta})$ to stand for the second, $\vec{\theta}$-dependent term, with the dropping of the first parameter-independent term understood.

Using eq.~\ref{eqn:logP}, the normalized trial probability distribution can be equivalently expressed in terms of a trial free energy surface $F_T(\vec{\xi}|\vec{\theta})$:
\begin{equation}
P_T(\vec{\xi}|\vec{\theta}) = \frac{e^{-F_T(\vec{\xi}|\vec{\theta})}}{\int e^{-F_T(\vec{\xi}'|\vec{\theta})} d\vec{\xi}'} \label{eqn:trialDist}
\end{equation}
If we set $W(\vec{x}) = W_{\mathrm{unbiased}}(\vec{x})$ to be the weighting function for our
unbiased reduced potential energy $u(\vec{x})$, and seek the trial
free energy surface in the unbiased state $F_T(\vec{\xi}|\vec{\theta}) = F(\vec{\xi}|\vec{\theta})$, the function to be minimized reduces to: 
\begin{eqnarray}
D_{\mathrm{KL}}(\vec{\theta}) &=& \int -P_E(\vec{\xi}|\{\vec{x}_n\}) \ln P_T(\vec{\xi}|\vec{\theta}) d\vec{\xi} \nonumber \\
               &=& \int P_E(\vec{\xi}|\{\vec{x}_n\}) F(\vec{\xi}|\vec{\theta}) d\vec{\xi} + \int P_E(\vec{\xi}) \ln \int e^{-F(\vec{\xi}'|\vec{\theta})} d\vec{\xi}' d\vec{\xi} \nonumber \\
               &=& \int P_E(\vec{\xi}|\{\vec{x}_n\}) F(\vec{\xi}|\vec{\theta}) d\vec{\xi} + \ln \int e^{-F(\vec{\xi}'|\vec{\theta})} d\vec{\xi}'  \nonumber \\
               &=& \sum_{n=1}^N W(\vec{x}_n) F(\vec{\xi}_n|\vec{\theta}) + \ln \int e^{-F(\vec{\xi}'|\vec{\theta})} d\vec{\xi}' \label{eq:kldiverge}
\end{eqnarray}
Between the 2nd and 3rd steps we integrate out the
$P_E(\vec{\xi}|\{\vec{x}_n\})$ term as $P_E(\vec{\xi}|\{\vec{x}_n\})$ is normalized, is
independent of the dummy variable $\vec{\xi}'$, and $\vec{\xi}_n =
\Phi(\vec{x}_n)$, and between the 3rd and 4th steps we employ eq.~\ref{eqn:expect} to estimate the expectation value over the data.  Minimization of eq.~\ref{eq:kldiverge} presents a prescription to adjust $\vec{\theta}$ to find the free energy surface $F(\vec{\xi}_n|\vec{\theta})$ which is the logarithm of
the closest distribution to the empirical delta function distribution calculated from
MBAR.  

Before proceeding to do so, it is instructive to make several observations about eq.~\ref{eq:kldiverge}. 
\begin{itemize}
 \item The biasing functions do not appear explicitly
   \textit{anywhere} in eq.~\ref{eq:kldiverge}. The biases appear only
   implicitly through the weights associated with samples from biased states. 
   One may therefore also carry out any
   other type of accelerated sampling, in addition to, or instead of
   biasing functions of the collective variable, as long as these
   simulations have a time-independent potential (they cannot involve
   adaptive biasing), and are included in the $K$ states for which
   MBAR reweighting is carried out and the weights $W(\vec{x}_n)$ are
   determined; the sum then is over \textit{all} points, collected in
   whatever simulation is used.
 \item The contribution $F(\vec{\theta}) = -\ln \int
  e^{-F(\vec{\xi}'|\vec{\theta})} d\vec{\xi}'$ is independent of the samples, and thus penalizes 
  free energy surfaces that are simply low everywhere. 
  \item Low free energy regions of the FES contribute more to 
  the integral $F(\vec{\theta}) = -\ln \int e^{-F(\vec{\xi}'|\vec{\theta})} d\vec{\xi}'$ than high free 
  energy regions. Accordingly, we should expect better estimates at the low values of $F$ (high probability states), but may sacrifice accuracy at large values of $F$ (low probability states). 
\end{itemize}

\textbf{C.2. Summed biased state Kullback-Leibler divergence.} We can measure
closeness to the KL divergence in a slightly different way, and try to
find a single function that minimizes the sum of KL divergences from
the $K$ empirical distribution functions observed at each biased
sample state to the trial function with the biased potential
added. The motivation for this ansatz is that it will force the trial function close to the
free energy surface in all regions the biased simulations have
high density and therefore good sampling. When summing over the $K$ different biased simulations, we elect to weight the KL divergence proportional to the number of samples
$N_k$ from that state. The motivation for this choice is that simulations with few samples should contribute less information than simulations with many. 
We will see that this assumption leads to particularly simple results.

Under these choices we define the sample-weighted sum of Kullback-Leibler divergences and function to be minimized as:
\begin{eqnarray}
\sum_{k=1}^{K} N_k D_{\mathrm{KL}}(\vec{\theta}) &=& \sum_{k=1}^K N_k \left(\sum_{n=1}^N W_k(\vec{x}_n) F_k(\vec{\xi}_n|\vec{\theta})\right. \nonumber \\
                              & & + \left. \ln \int e^{-F_{k}(\vec{\xi}'|\vec{\theta})} d\vec{\xi}'\right) \nonumber \\
                              &=& \sum_{k=1}^{K} N_k \left(\sum_{n=1}^N W_k(\vec{x}_n) \left(F(\vec{\xi}_n|\vec{\theta}) + b_k(\vec{\xi}_n)\right)\right. \nonumber \\
                              & & + \left. \ln \int e^{-F(\vec{\xi}'|\vec{\theta})-b_k(\vec{\xi}')} \right) d\vec{\xi}' \nonumber \\
                              &=& \sum_{n=1}^N \left(\sum_{k=1}^K N_k W_k(\vec{x}_n)\right) F(\vec{\xi}_n|\vec{\theta}) \nonumber \\
                              & & + \sum_{k=1}^{K} N_k \ln \int e^{-F(\vec{\xi}'|\vec{\theta})-b_k(\vec{\xi}')} d\vec{\xi}' \nonumber \\
                              &=& \sum_{n=1}^N F(\vec{\xi}_n|\vec{\theta}) \nonumber \\
                              & & + \sum_{k=1}^{K} N_k \ln \int e^{-F(\vec{\xi}'|\vec{\theta})-b_k(\vec{\xi}')} d\vec{\xi}'\label{eq:sumkldiverge}
\end{eqnarray}
where $F_k(\vec{\xi})$ is the free energy surface of the $k$th biased state, $F(\vec{\xi}_n)$ and $F_k(\vec{\xi}_n)$ are the values of $F$ and $F_k$ at $\Phi(\vec{x}_n) = \vec{\xi}_n$, $b_k(\vec{\xi}_n)$ is the value of the biasing potential associated with biased simulation $k$ at $\Phi(\vec{x}_n) = \vec{\xi}_n$, and $F_{k}(\vec{\xi}|\vec{\theta}) = F(\vec{\xi}|\vec{\theta}) + b_k(\vec{\xi})$. We note that in moving from the second to third line we dropped the term
$\sum_{k=1}^{K}\left(\sum_{n=1}^N W_k(\vec{x}_n) b_k(\vec{\xi}_n)\right)$ because it is
independent of the $\vec{\theta}$, and thus does not affect the
minimization, and in moving from the third to fourth line we appeal to the normalization condition for $W_k(\vec{x}_n)$ in eq.~\ref{eq:normal}. The latter operation eliminates the weights from each individual state, leaving as the first term in our final expression an unweighted sum over the trial functions at the empirical data points. The second term is a weighted sum over an integral over the trial functions and biasing potentials and contains significant contributions only where the biasing potential is low.
Large biasing potentials result in small contributions and essentially free variations of the trial function. However, as long as the trial
function has significant weight in one of the biasing functions, then
it will be constrained over that region of space.  In our
numerical tests discussed below, it appears that eq.~\ref{eq:sumkldiverge}
gives additional accuracy in the densely sampled regions
by sacrificing accuracy in the sparsely sampled regions, but provides superior global fits compared to those achieved by minimization of eq.~\ref{eq:kldiverge}. 

It is possible in many cases to include simulations performed with other accelerated sampling methods in addition to biasing in the collective variable, but unlike in this prototypical umbrella sampling case, the results are more complicated. We provide a preliminary analysis in the Appendix Section~\ref{sec:other_biasing}, but do not further analyze these combinations in this paper.

\textbf{C.3. Summed sampled biased state Kullback-Leibler divergence.} 
The final alternative we consider is to sum the KL divergences from the $K$ empirical distribution functions with the biased potential added as we do in the preceding section, but only using the $N_k$ actual samples from each biased state.  In this case, each weight will be simply $1/N_k$, as each of the $N_k$ samples will be equally weighted. We will continue to weight each state by the number of samples $N_k$ collected from the state, as states with more samples contribute proportionally more information to the KL divergence. Following a similar development to that which led to eq.~\ref{eq:sumkldiverge} and again dropping terms that are not dependent on $\vec{\theta}$ yields the expression to be minimized as: 
\begin{eqnarray}
\sum_{k=1}^{K} N_k D_{\mathrm{KL}}(\vec{\theta}) &=& \sum_{k=1}^K N_k \left(\sum_{n=1}^{N_k}\frac{1}{N_k}F_k(\vec{\xi}_n|\vec{\theta})\right. \nonumber \\
                              & & + \left. \ln \int e^{-F_{k}(\vec{\xi}'|\vec{\theta})} d\vec{\xi}'\right) \nonumber \\
                              &=& \sum_{k=1}^{K} N_k \left(\sum_{n=1}^{N_k} \frac{1}{N_k} \left(F(\vec{\xi}_n|\vec{\theta}) + b_k(\vec{\xi}_n)\right)\right. \nonumber \\
                              & & + \left. \ln \int e^{-F(\vec{\xi}'|\vec{\theta})-b_k(\vec{\xi'})} \right) d\vec{\xi}' \nonumber \\
                              &=& \sum_{k=1}^{K} \sum_{n=1}^{N_k} F(\vec{\xi}_n|\vec{\theta}) \nonumber \\
                              & & + \sum_{k=1}^K N_k \ln \int e^{-F(\vec{\xi}'|\vec{\theta})-b_k(\vec{\xi'})} d\vec{\xi}' \nonumber \\
                              &=& \sum_{n=1}^{N} F(\vec{\xi}_n|\vec{\theta}) \nonumber \\
                              & & + \sum_{k=1}^K N_k \ln \int e^{-F(\vec{\xi}'|\vec{\theta})-b_k(\vec{\xi'})} d\vec{\xi}'
\label{eq:weightedsimplesum}
\end{eqnarray}
Somewhat surprisingly, this result is exactly the same as eq.~\ref{eq:sumkldiverge}. This emerges due to the normalization condition for $W_k(\vec{x}_n)$ defined by eq.~\ref{eq:normal}. Accordingly, whether we
sum the contribution to the KL divergence of each sample over all
states using the MBAR weights, or simply sum the contribution of
each sample to its biased state, we will be minimizing the same
function, provided we weight by the number of samples $N_k$ from each distribution.

We could, in principle, also choose to sum over the $K$ KL divergences without weighting each biased distribution by $N_k$. Doing so and following the steps leading to eq.~\ref{eq:weightedsimplesum} yields the expression:  
\begin{eqnarray}
\sum_{k=1}^{K} D_{\mathrm{KL}}(\vec{\theta}) &=& \sum_{k=1}^{K}\frac{1}{N_k}\sum_{n=1}^{N_k} F(\vec{\xi}_n|\vec{\theta}) \nonumber \\
                              & & + \sum_{k=1}^K \ln \int e^{-F(\vec{\xi}'|\vec{\theta})-b_k(\vec{\xi}')} d\vec{\xi}'
\label{eq:simplesum}
\end{eqnarray}
which is less intuitively satisfying than eq.~\ref{eq:weightedsimplesum} since simulations conducted at a state point with small $N_k$ contribute equally to those with large $N_k$.  Likewise, if we follow the logic of eq.~\ref{eq:sumkldiverge} but employing equal weightings, we end up with a similarly unsatisfying result:   
\begin{eqnarray}
\sum_{k=1}^{K} D_{\mathrm{KL}}(\vec{\theta}) &=& \sum_{n=1}^{N} \left(\sum_{k=1}^K W_k(\vec{x}_n)\right) F(\vec{\xi}_n|\vec{\theta}) \nonumber \\
                              & & + \sum_{k=1}^K \ln \int e^{-F(\vec{\xi}'|\vec{\theta})-b_k(\vec{\xi}')} d\vec{\xi}'
\label{eq:simplesumkldiverge}
\end{eqnarray}
which is not only more complicated than eq.~\ref{eq:sumkldiverge}, but also differs (as numerical tests confirm) from eq.~\ref{eq:simplesum} unless all $N_k$ are equal, in which case $\sum_{k=1}^K W_k(\vec{x}_n)= K/N = 1/N_k$, and equality is restored. Due to these features, we will not pursue eq.~\ref{eq:simplesum} and eq.~\ref{eq:simplesumkldiverge} further.

\subsection{Likelihood as a measure of distance} \label{subsec:likelihood}

As an alternative to the Kullback-Leibler divergence, we can measure distances using likelihoods. Specifically, we can take our
trial probability distribution $P_T(\vec{\xi}|\vec{\theta})$ and
compute the \textit{likelihood} of one of our $N$ observations by
evaluating the $P_T$ associated with that observation. The observations taken together comprise our
data $D$.  Assuming the samples are independent and identically distributed (i.i.d.)\ observations, then we can calculate
the total likelihood as the  product of the individual likelihoods.  The trial probability distribution as a function of
$\theta$ that maximizes this likelihood will be the one closest to the
empirical distribution. In a similar manner to the KL divergence, we may  construct  this distribution in a number of ways. We shall show that the two choices we propose contain the same information as the KL divergence expressions, but offer greater interpretability and amenability to a Bayesian treatment.

\textbf{D.1. Product over unbiased state likelihoods.}  Perhaps the
simplest choice is to consider the joint likelihood of each weighted
sample in the unbiased state. In this case, since we can consider each
sample to be observed according to its weight $W(\vec{x}_n)N$ (the
expected number of counts at $\vec{x}_n$ given the empirical
distribution), then the overall likelihood as a function of
$\vec{\theta}$ is:
\begin{eqnarray}
 \ell(\vec{\theta}|\{\vec{x}_n\}) = \prod_{n=1}^N P_T(\vec{\xi}_n|\vec{\theta})^{W(\vec{x}_n)N} \label{eq:like1}
\end{eqnarray}
and the log likelihood is:
\begin{eqnarray}
 \ln \ell(\vec{\theta}|\{\vec{x}_n\}) &=& \sum_{n=1}^N NW(\vec{x}_n) \ln P_T(\vec{\xi}_n|\vec{\theta}) \nonumber \\
                 &=& \sum_{n=1}^N NW(\vec{x}_n) \left(-F(\vec{\xi}_n|\vec{\theta}) - \ln \int e^{-F(\vec{\xi}'|\vec{\theta}) d\vec{\xi}'}\right) \nonumber \\
                 &=& -N\sum_{n=1}^N W(\vec{x}_n) F(\vec{\xi}_n|\vec{\theta}) - N\ln \int e^{-F(\vec{\xi}'|\vec{\theta})} d\vec{\xi}' \nonumber \label{eq:likelihoodunbiased}
\end{eqnarray}
In going from the second to the third line, we employ normalization condition in eq.~\ref{eq:normal2}. As expected~\cite{Eguchi:JMA:2006}, we quickly verify that eq.~\ref{eq:likelihoodunbiased} is identical to eq.~\ref{eq:kldiverge} up to a factor of $(-N)$, so 
maximizing this log likelihood is the same as minimizing the unbiased state KL divergence.

\textbf{D.2. Product over biased state likelihoods.} We could also calculate the overall likelihood as the product of the likelihoods of the individual
samples in each of the biased simulations:
\begin{eqnarray}
 \ell(\vec{\theta}|\{\vec{x}_n\}) &=& \prod_{k=1}^{K} \prod_{n=1}^{N_k} P_T(\vec{\xi}_n|k,\vec{\theta}) \label{eq:like2}
\end{eqnarray}
where we have denoted the probability distribution resulting from the trial FES plus the $k$th bias as $P_T(\vec{\xi}_n|k,\vec{\theta})$. The corresponding log likelihood is:
\begin{eqnarray}
 \ln \ell(\vec{\theta}|\{\vec{x}_n\}) &=& \sum_{k=1}^K \sum_{n=1}^{N_k} \ln P_T(\vec{\xi}_n|k,\vec{\theta}) \nonumber \\
                 &=& \sum_{k=1}^K \sum_{n=1}^{N_k} \left(- F(\vec{\xi}_n|\vec{\theta}) - b_k(\vec{\xi}_n)\right. \nonumber\\
                 & & - \left.\ln \int e^{-F(\vec{\xi}'|\vec{\theta})-b_k(\vec{\xi}')} d\vec{\xi}'\right) \nonumber \\
                 &=& \sum_{k=1}^K \left(\sum_{n=1}^{N_k} -F(\vec{\xi}_n|\vec{\theta})\right.  \nonumber \\
                 & & \left.-N_k \ln \int e^{-F(\vec{\xi}'|\vec{\theta})-b_k(\vec{\xi}')} d\vec{\xi}'\right) \nonumber \\
                 &=& -\sum_{n=1}^{N} F(\vec{\xi}_n|\vec{\theta}) \nonumber \\
                 & & - \sum_{k=1}^K N_k \ln \int e^{-F(\vec{\xi}'|\vec{\theta})-b_k(\vec{\xi}')} d\vec{\xi}' \label{eq:likelihoodbiased}
\end{eqnarray}
where in going from the second to third line we drop the $b_k(\vec{\xi}_n)$ term as independent of
$\vec{\theta}$ and therefore irrelevant to the maximization. Eq.~\ref{eq:likelihoodbiased} is identical to eq.~\ref{eq:sumkldiverge} up to a minus sign, so maximizing the product of biased state likelihoods is equivalent to minimizing the summed biased KL divergence.

\textbf{D.3. Weighted product over biased state likelihoods.} We could try to construct a likelihood that was consistent with the KL divergence in eq.~\ref{eq:simplesum} by constructing a sum of KL divergences over each state weighted by the reciprocal of the number of samples in each state:
\begin{equation}
 \ell(\vec{\theta}|\{\vec{x}_n\}) = \prod_{k=1}^{K} \prod_{n=1}^{N_k} P_T(\vec{\xi}_n|k,\vec{\theta})^{\frac{1}{N_k}}, \label{eqn:likeWeight}
\end{equation}
for which the corresponding log likelihood is:
\begin{eqnarray}
 \ln \ell(\vec{\theta}|\{\vec{x}_n\}) &=& \sum_{k=1}^K \sum_{n=1}^{N_k} \frac{1}{N_k} \ln P_T(\vec{\xi}_n|k,\vec{\theta}) \nonumber \\
                 &=& \sum_{k=1}^K \frac{1}{N_k} \sum_{n=1}^{N_k} \left(- F(\vec{\xi}_n|\vec{\theta}) - b_k(\vec{\xi}_n)\right. \nonumber\\
                 & & - \left.\ln \int e^{-F(\vec{\xi}'|\vec{\theta})-b_k(\vec{\xi}')} d\vec{\xi}'\right) \nonumber \\
                 &=& \sum_{k=1}^K \frac{1}{N_k} \sum_{n=1}^{N_k} -F(\vec{\xi}_n|\vec{\theta})  \nonumber \\
                 & & - \sum_{k=1}^K \ln \int e^{-F(\vec{\xi}'|\vec{\theta})-b_k(\vec{\xi}')} d\vec{\xi}' \nonumber \\
                 &=& -\sum_{k=1}^K \frac{1}{N_k} \sum_{n=1}^{N_k} F(\vec{\xi}_n|\vec{\theta})  \nonumber \\
                 & & - \sum_{k=1}^K \ln \int e^{-F(\vec{\xi}'|\vec{\theta})-b_k(\vec{\xi}')} d\vec{\xi}' \label{eq:likelihooddirect}
\end{eqnarray}
Eq.~\ref{eq:likelihooddirect} is identical to eq.~\ref{eq:simplesum} up to a minus sign, and so maximizing the former is equivalent to minimizing the latter. However, as discussed above, there appears to be no real justification to weight samples in the manner expressed in eq.~\ref{eqn:likeWeight} and for this reason we do not advocate the use of this formulation.

\subsection{Least squares as a measure of distance}

Finally, we could choose to adopt a functional form, and then perform a least
squares fit to the empirical distribution or to the empirical
FES in order to define a distance between the distributions. Although seemingly quite a natural and straightforward approach, it does not give
rise to easily interpretable or implementable expressions. Accordingly, we defer an
analysis of the least squares approach to the Appendix Section~\ref{sec:least_squares} and do not pursue this further.

\subsection{How does vFEP fit into this framework?}

We now examine the correspondence of our development with the variational free energy profile (vFEP) approach developed by Lee and co-workers~\cite{Lee:JCTC:2013,Lee:JCTC:2014}.  

We first note the potential ambiguity within vFEP regarding the definition of the term ``window''. As described before, this could refer to a biasing potential, the data collected from a
simulation run with that biasing potential, or a region of
collective variable space within which a biased simulation has high
probability density. These are related, but not equivalent, concepts. 
In the present comparison with vFEP, we will assume ``window'' as used in the vFEP definition refers to a biasing
potential plus the data collected during simulations with that biasing
potential. 
Under this definition of ``window'', samples in the window are not included or excluded based on
the associated values of $\vec{\xi}$, only on the basis of biased simulation from which they were collected.

Using the original vFEP notation, $Z^{a} = \int e^{-F_{i,a}(\theta,x)} dx$ is the partition function of biased simulation $a$ and $F_{i,a}(\theta,x) = F_i(\theta,x) + W_a(x)$ is the biased trial partition function determined by parameters $\theta$ and collective variable $x$, where $W_a(x)$ is the biasing potential, and vectors in $x$ and $\theta$ are implicit. 
Since $W_a(x)$ is not a function of $\theta$ and does not affect the minimization, the log likelihood to be maximized with respect to the parameters $\theta$ of the trial function $F$ is:
\begin{eqnarray} 
\ln \ell(\theta) &=& \sum_a \left[-\ln Z^a - \frac{1}{N_a} \sum_{i=1}^N F_{i,a}(\theta,x_a)\right] \nonumber \\
          &=& \sum_a \left[ -\frac{1}{N_a} \sum_{i=1}^N F_i(\theta,x_a) - \ln \int_{\Gamma_a} e^{-F_{i,a}(\theta,x)} dx \right] \nonumber \\
\end{eqnarray}
To proceed, we must make two assumptions: (i) the substitution of $k$ as a label for biasing potential rather than $a$ as the label
of ``windows'', (ii) the recognition that
$\int_{\Gamma_a}$ should be either the same or approximately the same as $\int_\Gamma$, since
samples from biased potential will be mostly constrained to subsets of
$\Gamma$, but can in principle appear anywhere in $\Gamma$. In this case, we can translate vFEP into the terminology of the present paper. The window $a$ becomes the biased simulation $k$, $N_a$ becomes $N_k$, $x$ becomes $\vec{\xi}$, vectors are noted explicitly, and we obtain:
\begin{eqnarray} 
\ln \ell(\vec{\theta}|\{\vec{x}\}) &=& \sum_{k=1}^{K} \left[ -\frac{1}{N_k} \sum_{i=1}^{N_k} F(\vec{\xi}_i|\vec{\theta}) \right. \nonumber \\
                 & & \left. - \ln \int e^{-F(\vec{\xi}'|\vec{\theta})-b_k(\vec{\xi}')} d\vec{\xi}' \right]
\end{eqnarray}
This expression is identical to eq.~\ref{eq:likelihooddirect} and, up to a minus sign, eq.~\ref{eq:simplesum}. Accordingly, when viewed through the lens of the development presented in this paper---and with the previously mentioned
assumptions about the definitions of windows and range of integrals---vFEP would correspond to a particular choice of biased state weighting within a Kullback-Leibler divergence (eq.~\ref{eq:simplesum}) or likelihood formulation (eq.~\ref{eq:likelihooddirect}). 
As discussed above, this weighting of all simulations equally is problematic, since it puts equal weight on simulations regardless of 
how many samples they have. If the direct sum over biasing potentials is changed to one weighted by $N_k$, then it becomes eq.~\ref{eq:likelihoodbiased}, which both easier to work with and better justified, with umbrellas with larger numbers of samples having more weight.

\section{A Bayesian framework for FES estimation}

Equipped with the prescriptions to calculate likelihood of observations under
the different assumptions detailed in Section~\ref{subsec:likelihood}, we can switch to a Bayesian framework
to find distributions possessing desirable features of an analytical
form, continuity, and smoothness that is most consistent with
our understanding of $F(\vec{\xi})$. We note that our use of a likelihood formulation, which was shown to be fully consistent with the KL divergence framework, is crucial in opening the door to a Bayesian formulation.

At the first step in this framework, we take a candidate trial
distribution $P_T(\vec{\xi} | \vec{\theta})$ and optimize its
parameters $\vec{\theta}$ to form the maximum \textit{a posteriori}
probability (MAP) estimate of $P_T(\vec{\xi} | \vec{\theta})$. This estimate
maximizes the Bayes posterior probability of the trial distribution,
rather than simply the likelihood, given the collected (biased) samples
and MBAR estimates of the relative free energy differences
$\Delta f_{ij} = f_j - f_i$ between biased
states.

As we introduce our Bayesian formulation, we note that the free
energies emerging from the MBAR equations have no free parameters;
they are the only estimated normalizing constants satisfying the
self-consistent equations in eq.~\ref{equation:estimator-of-free-energies}. It is possible to employ a Bayesian
approach to free energy estimation by sampling of either the density of
states~\cite{Habeck:PRL:2012} or weights of each sample in the
unbiased state~\cite{Moradi:NC:2015}, allowing one to incorporate
additional priors about the simulations in addition to priors on the shape of
the free energy surface. However, since the free energy is defined
completely by the Boltzmann distribution, and since the MBAR equations
provide the lowest variance importance sampling estimator and are
asymptotically unbiased, then in the absence of other information
about the system, it is the simplest and least biased approach to employ MBAR estimates for $\{f_i\}$.

A difference from previous efforts is that we cast our approach
within a Bayesian framework that enables transparent incorporation
of Bayesian priors, Bayesian uncertainty quantification, and Bayesian
model selection about the functional form of the potential of mean
force. Although we do not do so here, this formalism also sets the
stage for adaptive sampling, in which regions of the probability
distribution containing the most uncertainty are identified for
additional biased sampling to optimally direct computational
resources. This is similar in spirit to, but would go beyond, the adaptive approach of Schofield, which presents an elegant means to alter the analytical representation of
the unbiased probability distribution to minimize uncertainty
\cite{Schofield:JPCB:2017}, to actually guiding the collection of
additional data to optimally reduce uncertainty in the estimated
distribution. 

We note that we follow a fairly standard Bayesian approach that can be
found in many textbooks and other resources; one excellent
presentation of Bayesian techniques in data analysis in general is
offered by Ref.~\citenum{Sivia::2006}. We also note that one of the
authors has previously presented a fully Bayesian treatment of WHAM in
Ref.~\citenum{Ferguson:JCC:2017} that goes into more detail about the
Bayesian aspects of parameter optimization as it applies to free energy surfaces.

Given the set of biased samples $\{\vec{x}_n\}$ and their collective variable mappings $\{\vec{\xi}_n\} = \{\Phi(\vec{x}_n)\}$ and the associated weights $W(\vec{x}_n$) in the (unbiased) thermodynamic state calculated from MBAR (eq.~\ref{eq:MBARweight}), we apply Bayes' theorem~\cite{Sivia::2006} to construct an expression for the posterior probability of the parameters $\vec{\theta}$ given the data $\{\vec{x}_n\}$, obtaining: 
\begin{align}
\mathcal{P}(\vec{\theta} | \{\vec{x}_n\}) &= \frac{\mathcal{P}(\{\vec{x}_n\} | \vec{\theta}) \mathcal{P}(\vec{\theta})}{P(\{\vec{x}_n\})} \label{eq:Bayes}
\end{align}
where $\mathcal{P}(\vec{\theta} |\{\vec{x}_n\})$ is the \textit{posterior
  probability} of the parameters $\vec{\theta}$ given the sampled
data, $\mathcal{P}(\{\vec{x}_n\} | \vec{\theta}) =  \ell(\vec{\theta}|\{\vec{x}_n\})$ is the earlier-defined
\textit{likelihood} specifying the probability of the collected
samples given the particular choice of parameters, $\mathcal{P}(\vec{\theta})$
is the \textit{prior probability} of the parameters before any data
have been collected, and $\mathcal{P}(\{\vec{x}_n\}) = \int \mathcal{P}(\{\vec{x}_n\} |
\vec{\theta}) \mathcal{P}(\vec{\theta}) d\vec{\theta}$ is the probability of
observing the samples that we did (the
\textit{evidence}), serves to normalize the posterior, and contains no dependence on the parameters $\vec{\theta}$. Importantly, the prior enables us to transparently encode any
prior beliefs or knowledge about the parameters into our analysis that can
serve to regularize and stabilize our estimation. 

The MAP estimate of the parameters follows from maximization of the log posterior is:
\begin{eqnarray} \label{eq:MAP}
\vec{\theta}^\mathrm{MAP}(\{\vec{x}_n)\} &=& \overset{\mathrm{argmax}}{\vec{\theta}} \ln \mathcal{P}(\vec{\theta} | \{\vec{x}_n\}) \nonumber \\
&=& \overset{\mathrm{argmax}}{\vec{\theta}} \left( \ln \mathcal{P}(\{\vec{x}_n\} | \vec{\theta}) + \ln \mathcal{P}(\vec{\theta}) \right) \nonumber \\
&=& \overset{\mathrm{argmax}}{\vec{\theta}} \left( \ln \ell(\vec{\theta}|\{\vec{x}_n\}) + \ln \mathcal{P}(\vec{\theta}) \right)
\end{eqnarray}
Exploiting our previous observation that maximizing a log likelihood is the same as minimizing the corresponding KL divergence from an empirical distribution~\cite{Eguchi:JMA:2006}, we can equivalently view maximization of the Bayes posterior (eq.~\ref{eq:MAP}) from a frequentist perspective as minimization of the Kullback-Leibler divergence or maximization of the log likelihood subject to regularization by the logarithm of the Bayes prior.  

To use eq.~\ref{eq:MAP} we need to adopt a form for the likelihood $\ell(\vec{\theta}|\{\vec{x}_n\})$ and prior $\mathcal{P}(\vec{\theta})$. The development in  Section~\ref{subsec:likelihood} suggests we adopt eq.~\ref{eq:like1} or~\ref{eq:like2} as candidates for the likelihood, where we explicitly assumed samples to be i.i.d.\ distributed. If the samples cannot be treated as i.i.d., then the counts $N$ or $N_k$ should be corrected by an inefficiency factor reflecting the presence of correlations in the sampling procedure~\cite{Gallicchio:JPCB:2005,Zhu:JCC:2012}. The simplest and most common choice for the prior is a uniform prior $\mathcal{P}(\vec{\theta})$ = 1. With no dependence on the model parameters $\vec{\theta}$, it drops out of the maximization in eq.~\ref{eq:MAP} and the MAP estimate $\vec{\theta}^\mathrm{MAP}$ becomes coincident with the maximum likelihood (ML) estimate $\vec{\theta}^\mathrm{ML}$:
\begin{equation} 
\vec{\theta}^\mathrm{ML}(\{\vec{\x}_n)\} = \overset{\mathrm{argmax}}{\vec{\theta}} \ln \ell(\vec{\theta}|\{\vec{x}_n\}). \label{eq:ML}
\end{equation}
In principle, arbitrary priors are admissible---even improper priors
that do not have a finite integral---provided the posterior is proper
(i.e., integrates to unity)~\cite{Gelman::2013}. In a Bayesian sense,
we use the prior to encode prior knowledge or belief about the
character of the probability distribution (such as smoothness of the
splines). In the frequentist sense, the prior serves to regularize the
probability estimate, providing bias-variance trade-off and
compensating for sparse data. In a practical sense, the appropriate prior to adopt depends on the form of the model selected $P_T(\vec{\xi} | \vec{\theta})$, the size and quality of the simulation data, and the degree of prior belief or understanding of the system.
Adopting the likelihood in eq.~\ref{eq:like1}, 
the maximization in eq.~\ref{eq:MAP} can be expressed as:
\begin{widetext}
\begin{align}
\vec{\theta}^{MAP}(\{\vec{x}_n\}) &= \overset{\mathrm{argmax}}{\vec{\theta}} \left[ -N\sum_{n=1}^N W(\vec{x}_n) F(\vec{\xi}_n|\vec{\theta}) - N\ln \int e^{-F(\vec{\xi}'|\vec{\theta})} d\vec{\xi}' + \ln \mathcal{P}(\vec{\theta}) \right] \nonumber \\
&= \overset{\mathrm{argmin}}{\vec{\theta}} \left[ N\sum_{n=1}^N W(\vec{x}_n) F(\vec{\xi}_n|\vec{\theta}) + N\ln \int e^{-F(\vec{\xi}'|\vec{\theta})} d\vec{\xi}' - \ln \mathcal{P}(\vec{\theta})\right] \nonumber \\
&= \overset{\mathrm{argmin}}{\vec{\theta}} \left[ N\sum_{n=1}^{N} W(\vec{x}_n) F(\vec{\xi}_n|\vec{\theta}) - \ln \mathcal{P}(\vec{\theta})\right] \quad \mathrm{s.t.} \quad \int_\Gamma e^{-F(\vec{\xi}_n|\vec{\theta})} d\vec{\xi} = 1 \label{eq:max1},
\end{align}
\end{widetext}
where in going from line 2 to 3 we have appealed to the proportionality relationship $P(\vec{\xi}|\vec{\theta}) \propto e^{-F(\vec{\xi}|\vec{\theta})}$ (eq.~\ref{eqn:logP}) and asserted that this distribution must be normalized.   

Adopting the product of likelihoods eq.~\ref{eq:like2} 
the maximization in eq.~\ref{eq:MAP} becomes:
\begin{widetext}
\begin{align}
\vec{\theta}^{MAP}(\{\vec{x}_n\}) &= \overset{\mathrm{argmax}}{\vec{\theta}} \left[ -\sum_{n=1}^{N} F(\vec{\xi}_n|\vec{\theta}) - \sum_{k=1}^K N_k \ln \int e^{-F(\vec{\xi}'|\vec{\theta})-b_k(\vec{\xi}')} d\vec{\xi}' + \ln \mathcal{P}(\vec{\theta}) \right] \nonumber \\
&=  \overset{\mathrm{argmin}}{\vec{\theta}} \left[ \sum_{n=1}^{N} F(\vec{\xi}_n|\vec{\theta}) + \sum_{k=1}^K N_k \ln \int e^{-F(\vec{\xi}'|\vec{\theta})-b_k(\vec{\xi}')} d\vec{\xi}' -  \ln \mathcal{P}(\vec{\theta})\right] \nonumber \\
&= \overset{\mathrm{argmin}}{\vec{\theta}} \left[ \sum_{n=1}^{N} F(\vec{\xi}_n|\vec{\theta}) - \ln \mathcal{P}(\vec{\theta})\right] \quad \mathrm{s.t.} \quad \int_\Gamma e^{-F(\vec{\xi}'|\vec{\theta})-b_k(\vec{\xi}')} d\vec{\xi}' = 1 \; \; \forall k \label{eq:max2}.
\end{align}
\end{widetext}

There are thus two approaches to find the MAP or ML estimate: an unconstrained minimization enforcing the normalization implicitly (second-to-last lines in eq.~\ref{eq:max1} and~\ref{eq:max2}), and a constrained minimization enforcing the normalization explicitly (last lines in eq.~\ref{eq:max1} and~\ref{eq:max2}). The constrained minimization versions of the above expressions can be solved using the method of Lagrange multipliers or through any other constrained optimization method such as the interior point method or sequential quadratic programming (SQP). The relative efficiency of the two approaches will depend on the details of software methods available as well as the particular forms of the biases and $F(\vec{\xi}_n|\vec{\theta})$.

\section{Model selection}

The Akaike information criterion (AIC) or Bayesian information criterion (BIC) provide a principled means to
discriminate between different possible choices for the Bayes prior
and the trial probability distribution, The AIC is defined as
\cite{Akaike:ITAC:1974}:
\begin{align}
AIC = 2k - 2 \ln  \ell(\vec{\theta}|\{\vec{x}_n\}), \label{eq:AIC}
\end{align}
where $k$ is the number of estimated parameters in the model. The BIC is defined as~\cite{Schwarz:AS:1978}:
\begin{align}
BIC = k\ln N - 2 \ln \ell(\vec{\theta}|\{\vec{x}_n\}), \label{eq:BIC}
\end{align}
where $N$ is the number of data points. If we compute $\vec{\theta} = \vec{\theta}^\mathrm{MAP}$ for a number of model choices $i$, we can use these parameter estimates to compute the set of AIC or BIC values $\{a_i\}$ for the candidate models. The model with the lowest $a_i$ is the single model that is best supported by the data.

A more sophisticated approach to model selection defines the smallest of the $\{a_i\}$ as $a_\mathrm{min}$, then assigns the relative likelihood of model $i$ as $r_i = e^{-\Delta_i/2} = e^{-(a_i - a_\mathrm{min})/2}$. The model weights follow from the normalized $r_i$ and provide the likelihood of model $i$~\cite{Schofield:JPCB:2017}:
\begin{align}
\omega_i = \frac{r_i}{\sum_k r_k} = \frac{e^{-\Delta_i/2}}{\sum_k e^{-\Delta_k/2}}.
\end{align}
Adopting a threshold $q$ = 0.05 (for example), the $\{r_i\}$ can be used to discard models from consideration and/or determine that there is insufficient evidence to choose one model over the other. The $\{\omega_i\}$ may also be used as weighting factors with which to construct a multi-model composed from the weighted sum of the predictions of each candidate model.

\section{Bayesian uncertainty quantification}

The $\vec{\theta} = \vec{\theta}^\mathrm{MAP}$ estimate represents the single best point estimate of the parameters of the trial distribution $P_{T}(\vec{\xi}|\vec{\theta})$ given the data
$\{\vec{x}_n\}$ and the prior $\mathcal{P}(\vec{\theta})$. Uncertainties around these point estimates may be approximated by analytical error expectations 
or through bootstrap estimation~\cite{Paliwal:JCTC:2011}. A fully Bayesian uncertainty estimate is defined by the distribution of $\vec{\theta}$ dictated by the Bayes posterior~\cite{Ferguson:JCC:2017}. Empirical samples of $\vec{\theta}$ from the Bayes posterior
may be generated using the Metropolis-Hastings algorithm. This Markov Chain Monte-Carlo (MCMC) approach generates a sequence of
parameter realizations that converges to the stationary distribution
of the Bayes posterior~\cite{Smith::2013}. Under this
approach we propose trial moves in $\vec{\theta}$ that are accepted or
rejected according to the Metropolis-Hastings acceptance criterion
\cite{Smith::2013,Hastings:B:1970}:
\begin{widetext}
\begin{align}
\alpha(\vec{\theta}^\nu | \vec{\theta}^\mu) &= \min \left[ \frac{\mathcal{P}(\vec{\theta}^\nu | \{\vec{x}_n\}) \cdot q(\vec{\theta}^\mu | \vec{\theta}^\nu)}{\mathcal{P}(\vec{\theta}^\mu | \{\vec{x}_n\}) \cdot q(\vec{\theta}^\nu | \vec{\theta}^\mu)}, 1 \right] \notag \\
&= \min \left[ \frac{\mathcal{P}(\{\vec{x}_n\} | \vec{\theta}^\nu) \cdot \mathcal{P}(\vec{\theta}^\nu) \cdot q(\vec{\theta}^\mu | \vec{\theta}^\nu)}{\mathcal{P}(\{\vec{x}_n\} | \vec{\theta}^\mu) \cdot \mathcal{P}(\vec{\theta}^\mu) \cdot q(\vec{\theta}^\nu | \vec{\theta}^\mu)}, 1 \right] \notag \\
&= \min \left[ \frac{\ell(\vec{\theta}^\nu|\{\vec{x}_n\}) \cdot \mathcal{P}(\vec{\theta}^\nu) \cdot q(\vec{\theta}^\mu | \vec{\theta}^\nu)}{\ell(\vec{\theta}^\mu|\{\vec{x}_n\}) \cdot \mathcal{P}(\vec{\theta}^\mu) \cdot q(\vec{\theta}^\nu | \vec{\theta}^\mu)}, 1 \right] \label{eqn:MH}
\end{align} 
\end{widetext}
where $\alpha(\vec{\theta}^\nu | \vec{\theta}^\mu)$ is the probability of accepting a trial move from parameter set $\vec{\theta}^\mu$ to parameter set $\vec{\theta}^\nu$, and $q(\vec{\theta}^\nu | \vec{\theta}^\mu)$ is the probability of proposing this trial move. We have invoked Bayes' Theorem (eq.~\ref{eq:Bayes}) in going from the first line to the second, and observe that (importantly) the evidence has canceled top and bottom. In going from the second line to the third, we employed the identity $\mathcal{P}(\{\vec{x}_n\} | \vec{\theta}) =  \ell(\vec{\theta}|\{\vec{x}_n\})$. In the event that symmetric trial move proposal probabilities are adopted such that $q(\vec{\theta}^\nu | \vec{\theta}^\mu) = q(\vec{\theta}^\mu | \vec{\theta}^\nu)$, the Metropolis-Hastings acceptance criterion reduces to the Metropolis criterion~\cite{Smith::2013,Metropolis:JCP:1953}:
\begin{align}
\alpha(\vec{\theta}^\nu | \vec{\theta}^\mu) &= \min \left[ \frac{\ell(\vec{\theta}^\nu|\{\vec{x}_n\}) \cdot \mathcal{P}(\vec{\theta}^\nu)}{\ell(\vec{\theta}^\mu|\{\vec{x}_n\}) \cdot \mathcal{P}(\vec{\theta}^\mu)}, 1 \right] \label{eq:Met}
\end{align}

We initialize the Markov chain from $\vec{\theta}^\mathrm{MAP}$ corresponding to the maximum of the Bayes posterior $\mathcal{P}(\vec{\theta} | \{\vec{\xi}_n\})$ and propose trial moves that maintain normalization $\int_\Gamma \mathcal{P}(\vec{\xi} | \vec{\theta}) d\vec{\xi} = 1$. By monitoring $\mathcal{L}(\vec{\theta} | \{\vec{x}_n\}) = \ln \left( \mathcal{P}(\{\vec{x}_n\} | \vec{\theta}) \mathcal{P}(\vec{\theta}) \right) = \ln \ell(\vec{\theta}|\{\vec{x}_n\}) + \ln \mathcal{P}(\vec{\theta})$---which is proportional to the Bayes posterior up to an additive constant  with no $\vec{\theta}$ dependence (eq.~\ref{eq:Bayes})---we can determine that the Markov chain has converged when $\mathcal{L}(\vec{\theta} | \{\vec{x}_n\})$ plateaus to fluctuate around a stable mean. At this point we may harvest realizations of $\vec{\theta}$ distributed according to the Bayes posterior. Using these parameter realizations, we can construct realizations of $P_T(\vec{\xi}|\vec{\theta})$ to quantify the uncertainties in this estimated distribution. 

\section{Example: Umbrella sampling of protein sidechain torsion within binding cavity}

\begin{figure}[h]
\includegraphics[width=0.6\columnwidth]{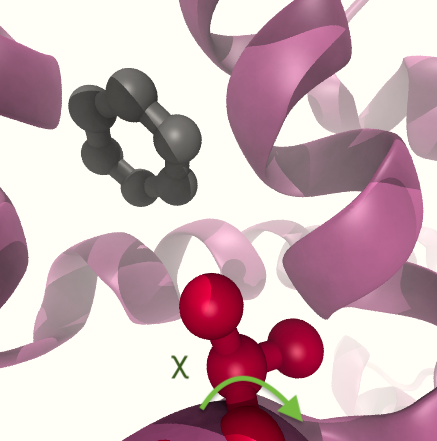}
\caption{$\chi$ torsion angle in Lys111 of L99a T4 lysozyme around which the free energy profile is calculated using umbrella sampling\label{fig:lyspic}.}
\end{figure}

As an illustrative example, we consider the application of our mathematical framework to compute a 1D FES from an
umbrella sampling simulation. Code
implementing these methods can be found publicly available in the
\texttt{pymbar4} branch of \texttt{pymbar} (located at \texttt{http://github.com/choderalab/pymbar}), in the
script \\\texttt{examples/umbrella-sampling/\\umbrella-sampling-advanced-fes.py}. The data is from an umbrella sampling
simulation for the $\chi$ torsion of a valine sidechain in lysozyme
L99A with benzene bound in the cavity~\cite{Mobley:JMB:2007} (fig.~\ref{fig:lyspic}). 

We analyze data from 26 biased simulations employing umbrella
potentials at a range of dihedral values with harmonic biasing
constants of between 100 and 400 kJ/mol/nm$^2$. A 100 ps simulation was
carried out under each umbrella potential with angles and energies
saved every 0.2 ps for a total of 500 samples at each state. The data
was analyzed for correlations, and approximately every other data
point is taken (exact frequency varying with state) for a total of
7446 data points, ranging from 42 to 410 points per umbrella.

We examine the histogram approach (with 30 bins, a number chosen to be
visually clear---the number of bins can be chosen completely
independently of the number of umbrella simulations run), and the
kernel density approximation with a Gaussian kernel, with bandwidth
parameter half of the bin size, in this case, $\frac{1}{2} \times
360/30 = 6$ degrees.  We also look at parameterized splines as our
representation; in this case, using B-splines with varying numbers of
knots placed uniformly, using cubic splines in this example; the
theory is independent of these particular choices of spline.

We note that one could use splines to fit to either the FES $F(\vec{\xi}|\vec{\theta})$ or the probability distribution
$P(\vec{\xi}|\vec{\theta})$. However, we find that it becomes difficult to satisfy the non-negativity condition of $P(\vec{\xi}|\vec{\theta})$ when using standard spline implementations, and that large changes in FES propagate exponentially to the probability distribution making it challenging to fit stably and robustly. For numerical stability, we therefore recommend using splines to approximate $F(\vec{\xi}|\vec{\theta})$ rather than $P(\vec{\xi}|\vec{\theta})$.

We examine the parameterized spline representations emerging from the optimizations defined by the expressions in eq.~\ref{eq:max1}---corresponding to the unbiased state likelihood in eq.~\ref{eq:like1}, log likelihood in eq.~\ref{eq:likelihoodunbiased}, and KL divergence in eq.~\ref{eq:kldiverge}---and eq.~\ref{eq:max2}---corresponding to the product of biased states likelihood in eq.~\ref{eq:like2}, log likelihood in eq.~\ref{eq:likelihoodbiased}, and KL divergence in eq.~\ref{eq:sumkldiverge}. We will refer to the first as the ``unbiased state likelihood'', and the second as the ``biased states likelihood,'' as it combines samples from all biased states.

\begin{figure}[h]
\includegraphics[width=0.5\textwidth]{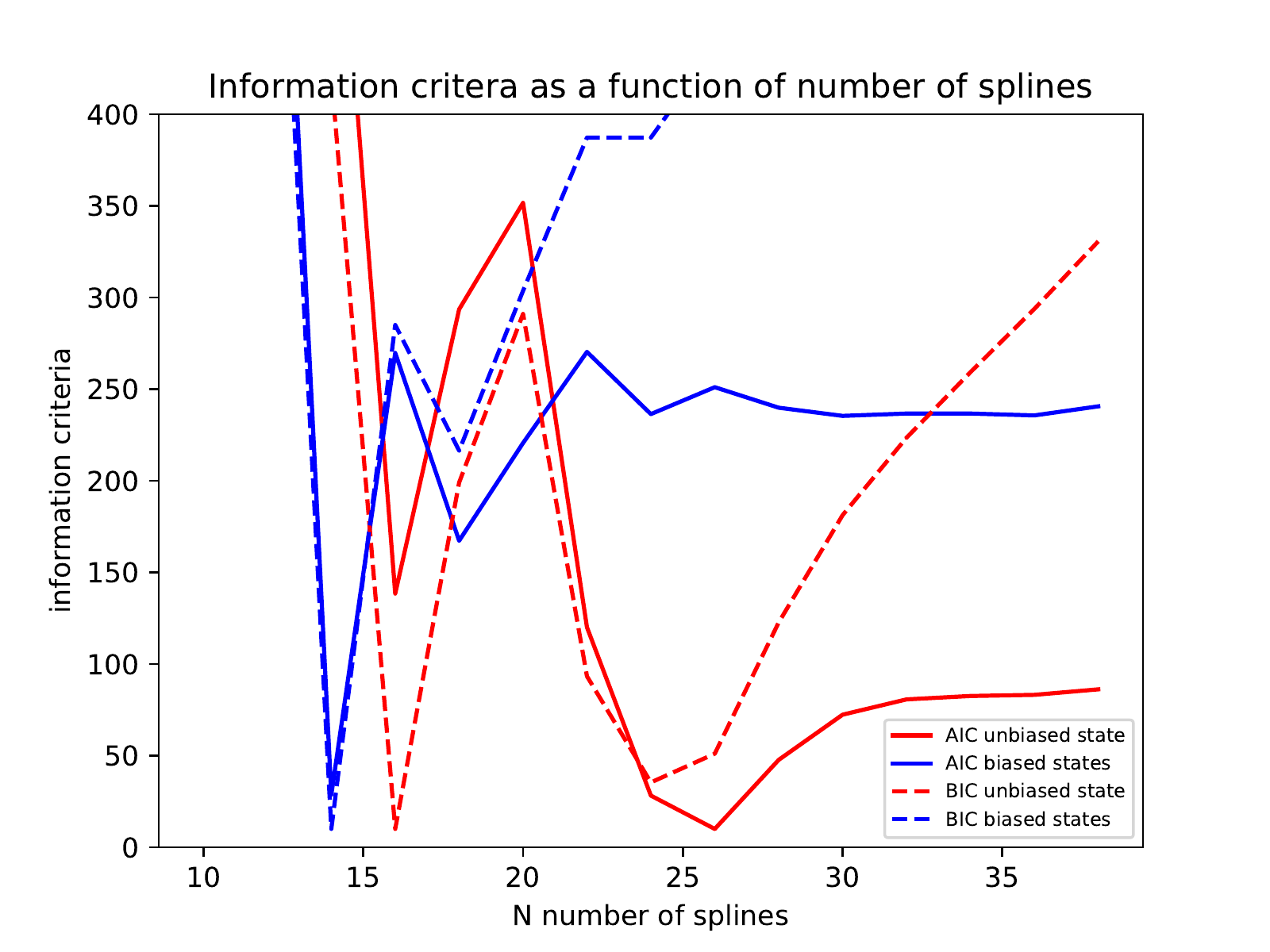}
\caption{AIC (solid) and BIC (dotted) for splines maximizing FES
  likelihoods for the unbiased state estimator (red,
  eq.~\ref{eq:likelihoodunbiased}) and biased states estimator (blue,
  eq.~\ref{eq:likelihoodbiased}) as a function of the number of spline
  knots, referenced from the minimum of each method.  Although the
  curves are noisy and nonmonotonic, they provide a
  useful guide towards choosing optimal numbers of parameters for
  models, as can be seen by comparison to
  Fig.~\protect{\ref{fig:compare_pmf}}.~\label{fig:IC}}
\end{figure}

Efficient optimization of these expressions requires calculating the
gradient and potentially the Hessians. The use of
B-splines, which construct the spline in terms of local basis function, makes this
calculation relatively efficient, as detailed in the
Appendix Section~\ref{sec:derivatives}. For simplicity, we elect to use a uniform distribution of spline knot locations over the domain, but these could be adaptively situated by optimizing their locations to maximize the MAP as proposed by Schofield~\cite{Schofield:JPCB:2017}. 

For the Bayes prior, when we compute the full posterior, rather than just the likelihood, we adopt a unnormalized Gaussian prior on the difference between successive spline knot values:
 \begin{align}
\mathcal{P}(\vec{\theta}) = \prod_{c=1}^{C-1} e^{-\alpha (\theta_c-\theta_{c+1})^2} \label{eq:smooth_prior}
 \end{align}
where $\alpha$ is a hyperparameter that controls the degree of smoothing regularization imposed upon the trial distribution. Selecting $\alpha$ = 0 corresponds to a uniform prior that drops out of the maximization and $\vec{\theta}^\mathrm{MAP} = \vec{\theta}^\mathrm{ML}$. Selecting $\alpha$ $>$ 0 favors smoother splines with less variation from knot to knot. 
We examine the effect of priors governed by choice of $\alpha$, where $\alpha = k/n$, where $n$ is the number of spline knots, for some constant $k$. Uncertainties are estimated by MCMC sampling of the Bayes posterior using the Metropolis-Hastings algorithm and acceptance criteria (eq.~\ref{eqn:MH}).

The time limiting factor, both for optimizations and MCMC sampling of
the posterior, is the numerical quadrature of the integral $\int
P_T(\vec{\xi} | \vec{\theta}) d\vec{\xi}$.  For the log likelihoods
from the unweighted state (eq.~\ref{eq:likelihoodunbiased}), the
integral enforcing the normalization of $P_T$ is only carried out over
the unbiased trial function, whereas for approaches considering all
states (eq.~\ref{eq:likelihoodbiased}), the integral is carried out
over all $K$ trial functions with biases and is thus roughly $K$ times slower.

The AIC and BIC allow us to select the number of spline knots best supported by the data. We
plot in fig.~\ref{fig:IC} the AIC (eq.~\ref{eq:AIC}) and BIC
(eq.~\ref{eq:BIC}) for the unbiased state likelihood and
biased states likelihood choices. In the unbiased state case, the AIC exhibits a local minimum at 16 knots and a global minimum at 26, whereas the BIC---which penalizes excessive parameters more strongly than the AIC---possesses a local minimum at 24 knots and a global minimum at 16. In the biased states case, the AIC and BIC both exhibit clear global minima at 14 knots.

\begin{figure}[h]
\begin{center}
\begin{subfigure}[t]{\columnwidth}
\centering
\includegraphics[width=0.95\textwidth]{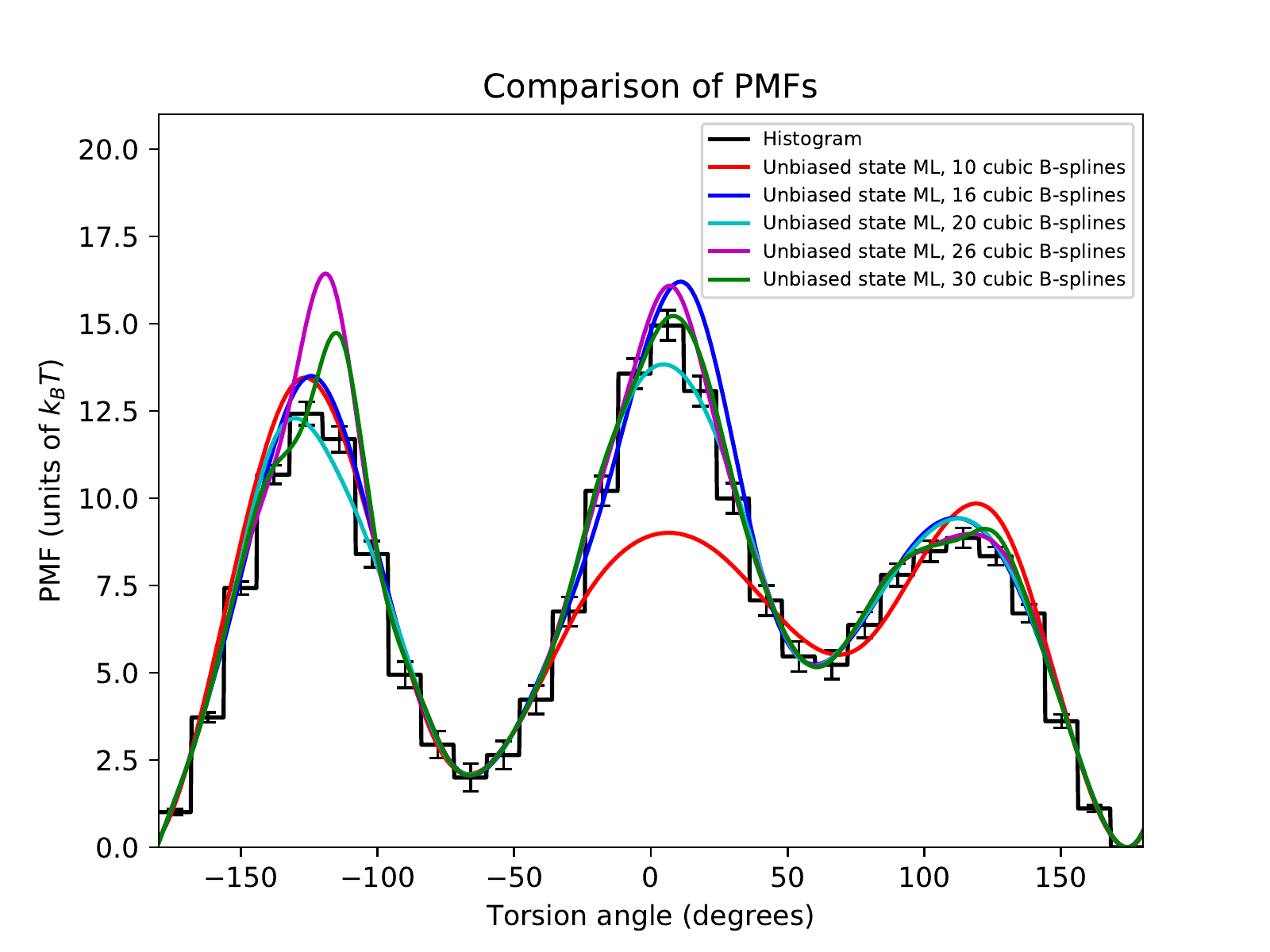}
\caption{\label{fig:pmf_unbiased}}
\end{subfigure}
\begin{subfigure}[t]{\columnwidth}
\centering
\includegraphics[width=0.95\textwidth]{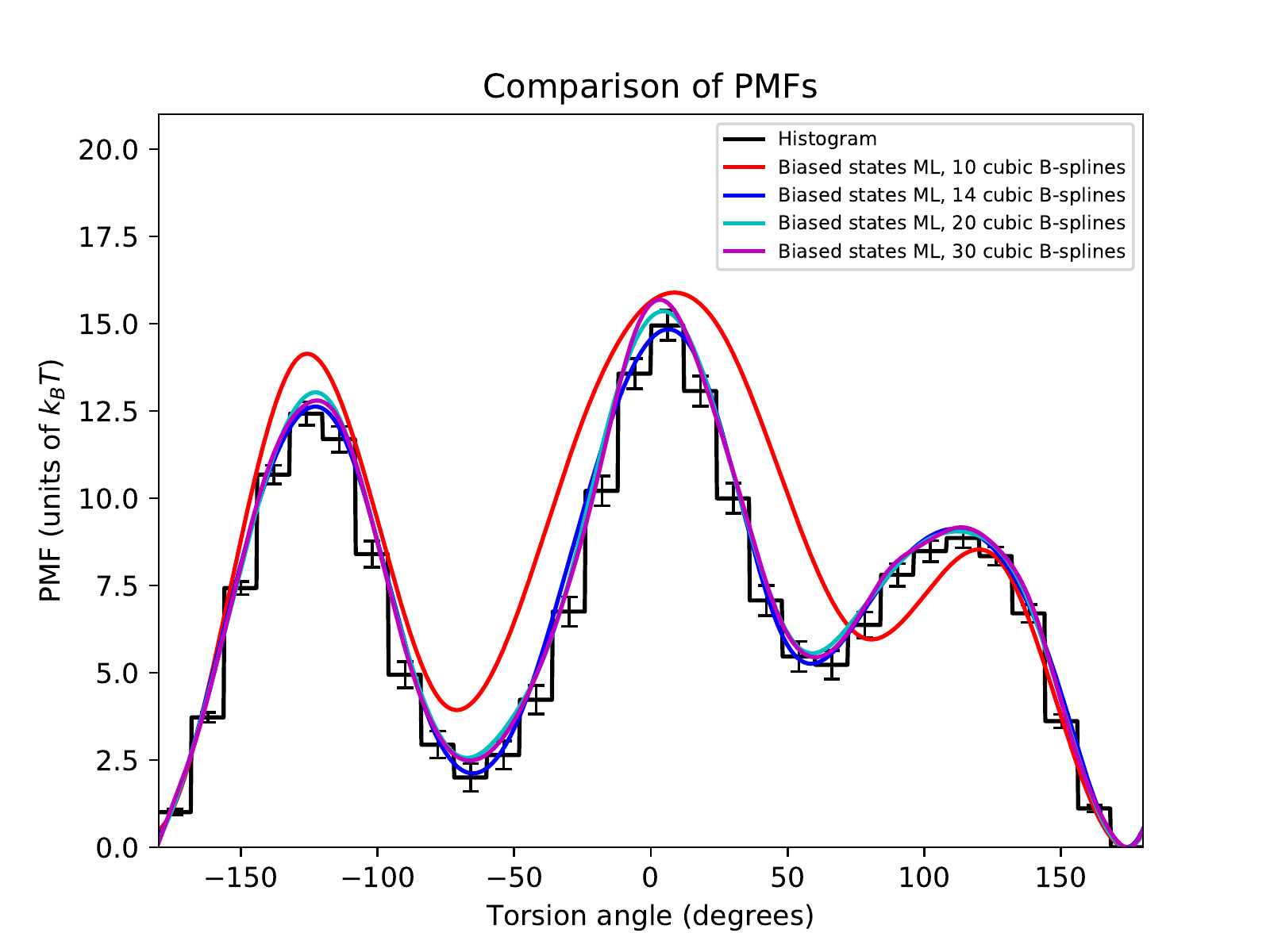}
\caption{\label{fig:pmf_biased}}
\end{subfigure}
\caption{Splines maximizing the (a) unbiased state
  likelihood (eq.~\ref{eq:likelihoodunbiased} or eq.~\ref{eq:max1}
  with uniform prior) and (b) biased states
  likelihood (eq.~\ref{eq:likelihoodbiased} or eq.~\ref{eq:max2}
  with uniform prior) as a function of the number of spline
  knots, with a histogram (black) as a reference. Knot numbers identified as optimal by both AIC
  and BIC appear to be good fits compared to other numbers of splines
  that under- or overfit the curve~\label{fig:compare_pmf}.} 
\end{center}
\end{figure}

We can see how the behavior of FES changes as a function of the number of knots and how the AIC and BIC help select optimal knot numbers in fig.~\ref{fig:compare_pmf}. In this figure, we plot maximum likelihood FES under
the unbiased state likelihood (eq.~\ref{eq:max1}, in
fig.~\ref{fig:pmf_unbiased}) and biased states likelihood
(eq.~\ref{eq:max2}, in fig.~\ref{fig:pmf_biased}) as a function of the
number of spline knots, along with the histogram estimate equipped with
uncertainties generated from error propagation from the weights via
MBAR~\cite{Shirts:JCP:2008}. As expected, higher numbers of knots
provide improved fitting, but overfitting becomes clear for larger
numbers of knots, especially in the case of fits using the unbiased
state likelihood. However, model complexities corresponding to AIC/BIC
minima fit the data relatively well in both cases. We note that the
unbiased state FES fits in fig.~\ref{fig:pmf_unbiased}, even for the
10-knot spline, are tightly grouped at the various FES minima, but they vary
significantly at the maxima, as there are less constraints on the
maxima than the minima using this approach.  In contrast, all fits with sufficient
functional flexibility (more than 10 spline knots) using the biased
states approach agree relatively well across the entire range of the
FES (fig.~\ref{fig:pmf_biased}), even with as few as 14 spline knots, the value corresponding to
the minimum of both AIC and BIC for the biased states likelihood.

\begin{figure}[h]
\includegraphics[width=0.5\textwidth]{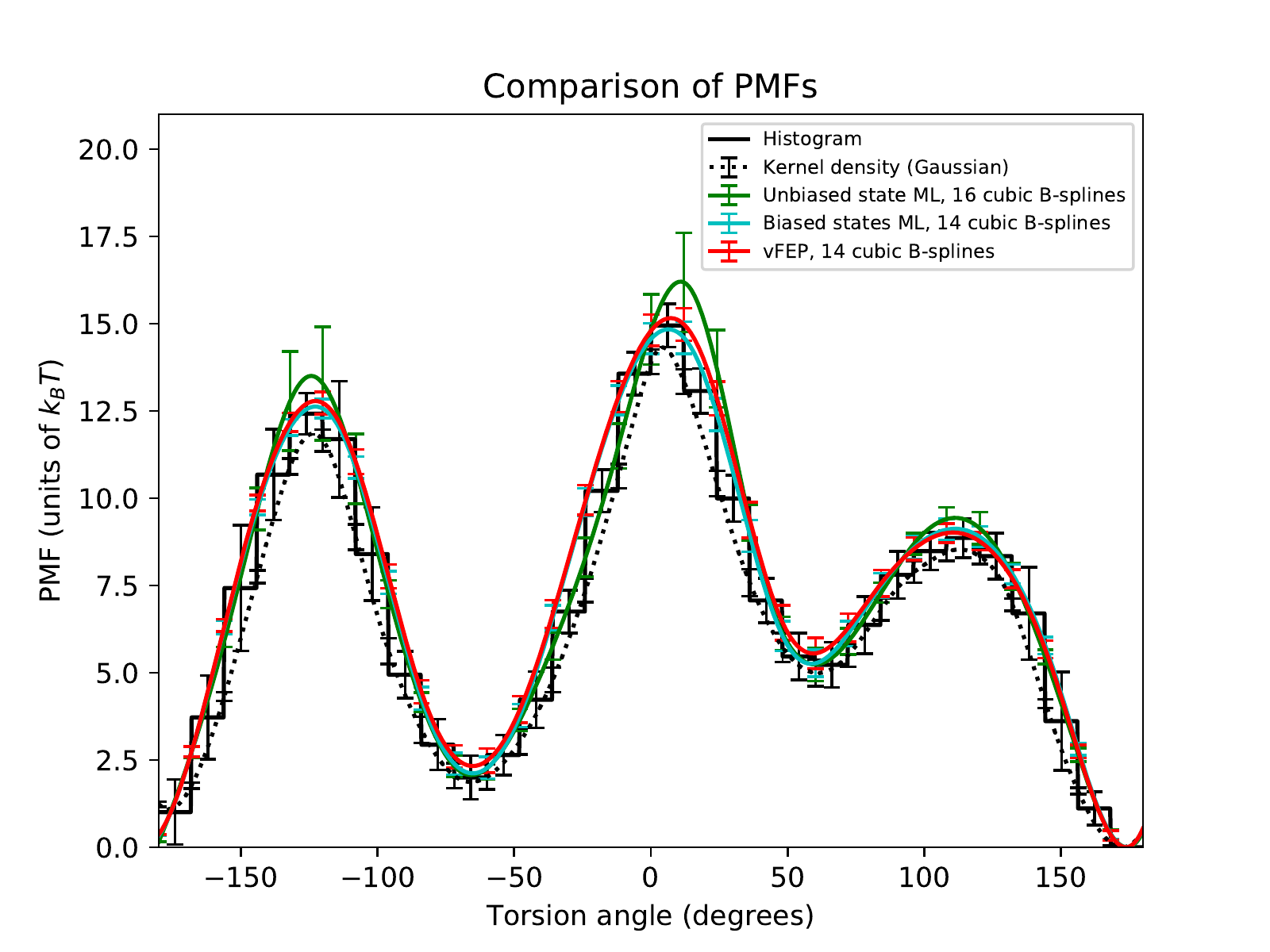}
\caption{Comparison of methods including with bootstrap uncertainty estimates. The number of splines employed in each method was selected according to the AIC / BIC analysis in fig.~\protect{\ref{fig:IC}}. The same number of spline knots is used for vFEP as for the biased states estimator. The histogram employs 30 bins and Gaussian kernels with $\sigma$ = 6$^\circ$. Uncertainties are estimated by bootstrap resampling with $n$ = 40. We observe that error bars are significantly greater at the barriers for the FES maximizing the likelihood in eq.~\ref{eq:likelihoodunbiased} than maximizing the likelihood in eq.~\ref{eq:likelihoodbiased}, which has very low uncertainty throughout the entire range of
  values. Histogram uncertainties are moderately large over the entire range.  \label{fig:withbars}}
\end{figure}

Adding bootstrapped uncertainty estimates to the FES help better show the relationship between the methods and their strengths and weaknesses. We present in Fig.~\ref{fig:withbars} a
comparison of the histogram (with 30 bins), kernel density
approximation (with Gaussian kernels with $\sigma$ of 6$^\circ$),
unbiased state likelihood and biased states likelihood splines
employing the AIC/BIC optimal number of knots, and vFEP (using with
the same number of splines as the biased states likelihood case).
Uncertainties all estimates are estimated from an ensemble of 40
bootstrap samples from each of the umbrellas.  All methods give
relatively similar results, which is to be expected with a well-sampled system and careful selection of parameters. 
In particular, the FES calculated using vFEP (subject to the assumptions discussed earlier in the text) is close to the biased states likelihood approximation. This result is expected because the two approaches coincide in the limit of equal numbers of uncorrelated samples per state.

\begin{figure*}[h]
\begin{center}
\begin{subfigure}[t]{0.495\textwidth}
\centering
\includegraphics[width=\linewidth]{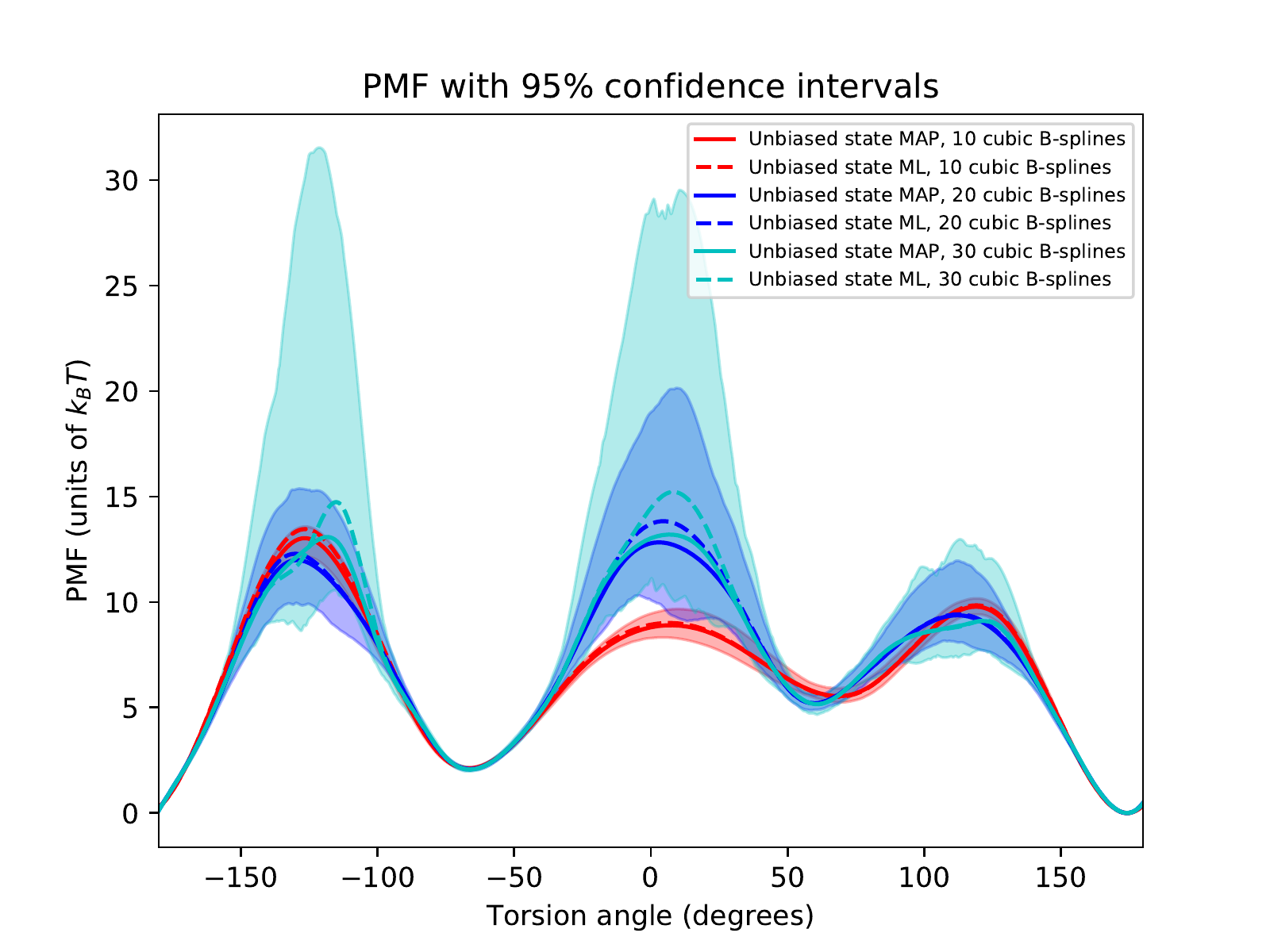}
\caption{\label{fig:mcmc_unbiaseda}}
\end{subfigure}
\begin{subfigure}[t]{0.495\textwidth}
\centering
\includegraphics[width=\linewidth]{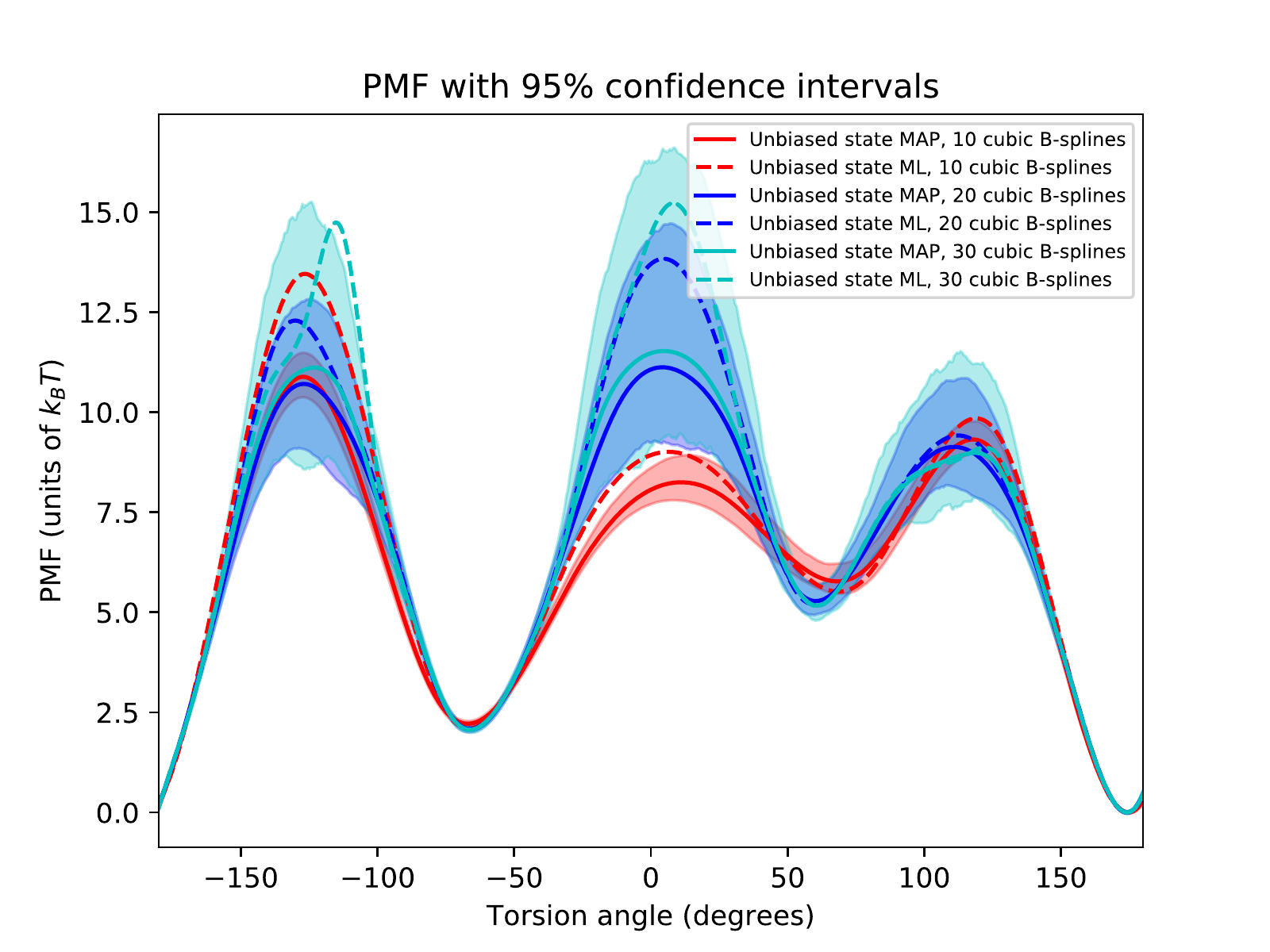}
\caption{\label{fig:mcmc_unbiasedb}}
\end{subfigure}

\begin{subfigure}[t]{0.495\textwidth}
\centering
\includegraphics[width=\linewidth]{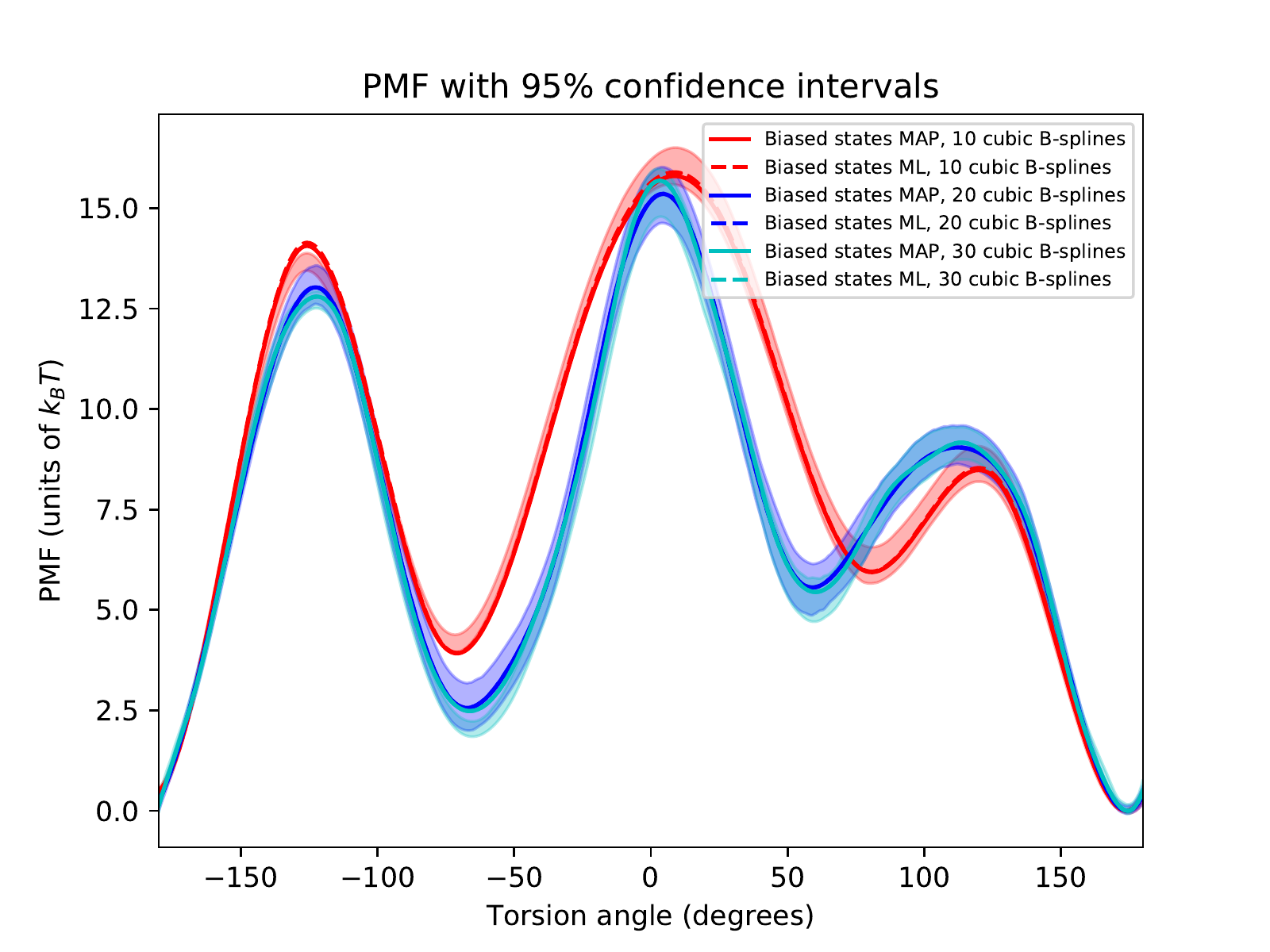}
\caption{\label{fig:mcmc_biasedc}}
\end{subfigure}
\begin{subfigure}[t]{0.495\textwidth}
\centering
\includegraphics[width=\linewidth]{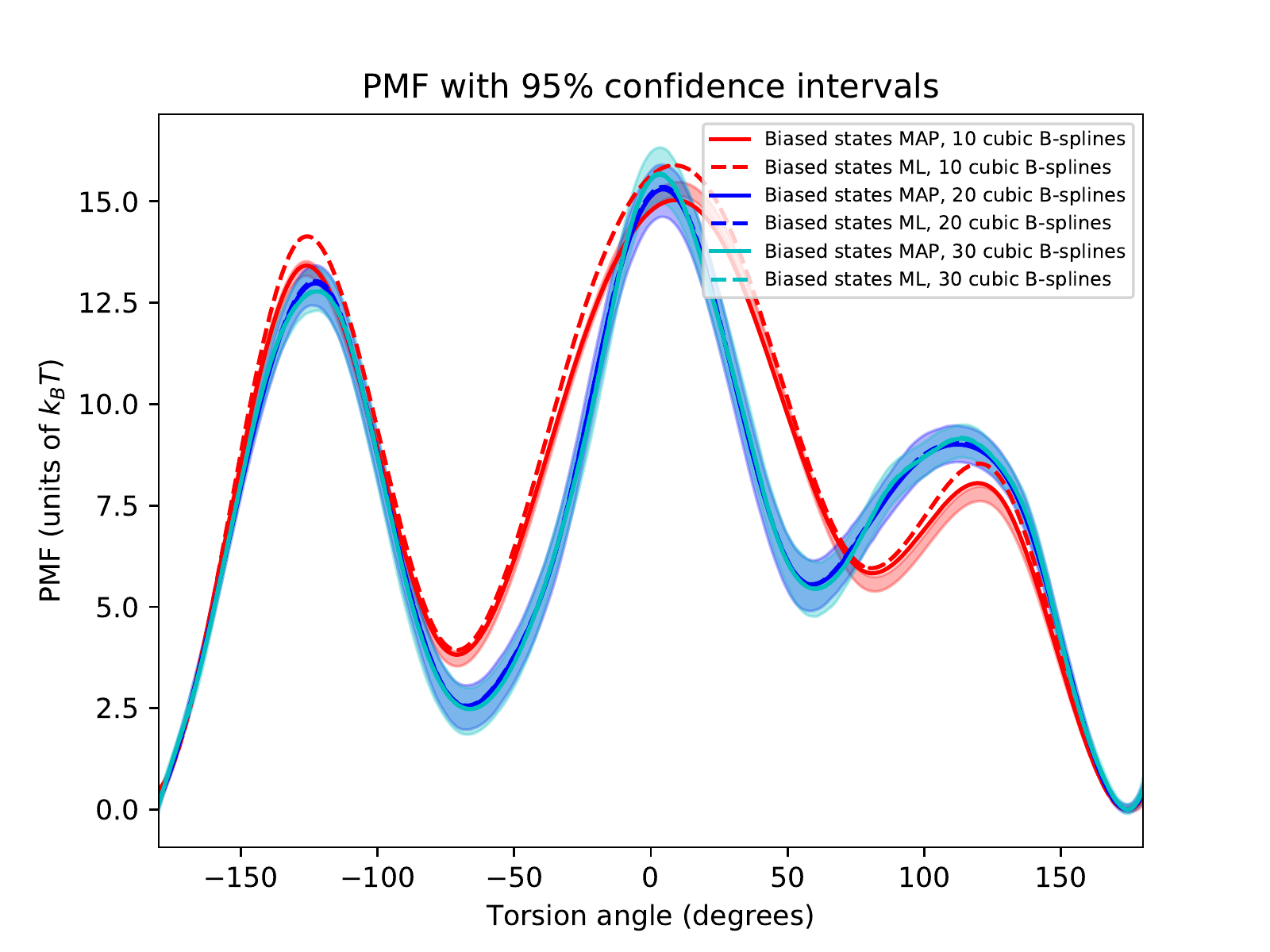}
\caption{\label{fig:mcmc_biasedd}}
\end{subfigure}
\caption{\label{fig:mcmc} Comparison of MAP estimates as a function of the number of spline knots with uncertainty estimates and a Gaussian prior (eq.~\ref{eq:smooth_prior}) with (a, c) $\alpha=0.1/n$ and (b, d) $\alpha=1/n$, where $n$ is the number of spline knots. We illustrate the MAP distributions for (a,b) the unbiased state likelihood (eq.~\ref{eq:max1}) and (b,d) biased states likelihood (eq.~\ref{eq:max2}). The shading represents the 95\% confidence intervals in the MAP estimate evaluated at each spline knot by MCMC sampling of the posterior, and the dashed line represents the ML solution. The MAP and ML solutions are coincident for $\alpha=0$. The choice $\alpha=0.1/n$ results in only minor differences between the ML and MAP solutions, whereas $\alpha=1/n$ results in a visible difference between the two curves. In the biased states formulation (figs.~\ref{fig:mcmc_biasedc} and ~\ref{fig:mcmc_biasedd}), the uncertainties are approximately constant across the range of the FES, whereas under the unbiased state formulation (figs.~\ref{fig:mcmc_unbiaseda} and~\ref{fig:mcmc_unbiasedb}), the uncertainty is largest at the high free energy regions where the likelihood function is least constrained.}
\end{center}
\end{figure*}

In fig.~\ref{fig:mcmc} we demonstrate the utility of fully Bayesian uncertainty
quantification. Uncertainties in the MAP splines are computed from 50,000
(for biased states posteriors, which is slower) and 200,000 (for unbiased
state posteriors) steps of MCMC sampling from the Bayes
posterior. Uncertainties represent the 95\% confidence intervals at
each spline knot. In both cases, we show results for 10, 20, and 30
splines for two different Gaussian priors (eq.~\ref{eq:smooth_prior}): (i)
$\alpha=0.1/n$ in fig.~\ref{fig:mcmc_unbiaseda} and fig.~\ref{fig:mcmc_biasedc}, where $n$ is the
number of spline knots, and (ii) $\alpha=1/n$ in fig.~\ref{fig:mcmc_unbiasedb} and
fig.~\ref{fig:mcmc_biasedd}. We recall that larger values of $\alpha$ impose a stronger influence of the smoothing prior and are expected to result in smoother posterior distributions. The choice of $\alpha=0.1/n$ produces very minor differences
between the ML and MAP curves (fig.~\ref{fig:mcmc_unbiaseda} and~\ref{fig:mcmc_biasedc}), whereas $\alpha=1/n$
results in a visibly apparent difference between the two curves (fig.~\ref{fig:mcmc_unbiasedb} and~\ref{fig:mcmc_biasedd}). We see that under
the biased states formulation (figs.~\ref{fig:mcmc_biasedc} and
~\ref{fig:mcmc_biasedd}), uncertainties are relatively low and constant across the full range of the FES, whereas in the unbiased state formulation (figs.~\ref{fig:mcmc_unbiaseda} and
~\ref{fig:mcmc_unbiasedb}), the uncertainties are largest at the high free energy regions where the likelihood function is least constrained (cf.~eq.~\ref{eq:max1}). Under the unbiased state formulation, the stronger smoothing prior with $\alpha=1/n$
(fig.~\ref{fig:mcmc_unbiasedb}) is valuable in reducing the size of the confidence intervals at the peaks of the FES (note the larger y-axis range in fig.~\ref{fig:mcmc_unbiaseda} required to accommodate the large uncertainty envelopes). We note that due to the
significant freedom in the 30-knot splines, MCMC sampling of the
probability nearly diverges in fig.~\ref{fig:mcmc_unbiaseda} with
$\alpha=0.1/n$.  In contrast, the biased states formulation provides
more constraints across the entire FES (cf.~eq.~\ref{eq:max2}), and the MCMC error bounds are
smaller over the entire range of the FES for both choices of $\alpha$
(fig.~\ref{fig:mcmc_biasedc} and ~\ref{fig:mcmc_biasedd}).

\section{Conclusions}

In this article, we have presented a Bayesian formalism to compute
free energy surfaces from the empirical distributions generated by
biased sampling, most simply with umbrella sampling in the collective
variables of interest, but capable of incorporating other accelerated
sampling methods as well. Within this formalism, we avoid any arbitrary choice
of histogram in either the definition of the FES or the calculation of
the weights, and provide clear and explicit criteria to decide which
continuous free energy surfaces are most consistent with the biased
sampling data. The choice and optimization of the representation of
the continuous FES is completely decoupled from the choice of biasing
functions and calculation of the relative free energies between the
biased simulations.  Biasing functions of the collective variables can
be chosen, with freedom of the biasing functional form, to give
appropriate sampling along the collective variables of interest, and
the samples and their associated Boltzmann weights are used to
construct the FES. The The Bayesian formalism allows us to choose the
FES that is sufficiently close to the empirical distribution of the
samples we have collected, and explicitly include any prior
information that we include by our choice of representation of our FES
functional form.  Our development also clearly demonstrates the
equivalence of the likelihood-based Bayesian formulation and
Kullback-Leibler-based frequentist formulation.

We find that the maximum likelihood calculated only from the unbiased
state (eqs.~\ref{eq:likelihoodunbiased} and~\ref{eq:kldiverge}) has a tendency to underestimate the 
free energy barriers in the collective variable.  The product of likelihoods from
all the unweighted samples collected from each biased state, weighted
by the number of samples collected from each biased state
(eqs.~\ref{eq:likelihoodbiased} and~\ref{eq:sumkldiverge}), has
much better overall performance over the entire FES range.
Surprisingly, this likelihood is exactly equal to the likelihood
generated from the product over all states of the reweighted
contribution of \textit{all} samples to each biased state state, again
weighted by the number of samples collected from each state
(cf.~eqs.~\ref{eq:sumkldiverge} and~\ref{eq:weightedsimplesum}).
 
We can then take these likelihoods and directly incorporate them into a
Bayesian inference framework.  Priors on the parameters of the FES can
then be chosen using whatever criteria is most appropriate; in this
study we considered a Gaussian prior enforcing smoothness, 
but the selection can be made based on any user-defined criteria, such as tethering free energies
to particular values or enforcing similarity to previously estimated
distributions. We can then use MCMC sampling of the posterior
of the FES curves to perform uncertainty quantification for arbitrary choices of prior.

We demonstrate our approach in an application to calculation of the FES for
the leucine rotation in the L99A mutant of T4 lysozyme. The unbiased state likelihood has some
clear failures in that it insufficiently constrains the FES at the
highest points.  This failure shows up in multiple ways.  When
computing bootstrap uncertainties, the unbiased states approach has
very high uncertainty in the barriers. With MCMC sampling, the
issues become even clearer, with significant fluctuation in the
parameters at the barriers unless a relatively severe prior is imposed. The biased states likelihood, however, behaves much more stably, with a well-constrained FES over the
entire range, even under weak priors.

Code implementing this approach is distributed in \texttt{pymbar},
where the previous free energy surface functionality, using
histograms to represent the FES, is replaced with a more comprehensive
\texttt{pymbar.FES} module implementing the formalism presented in this paper.

The Bayesian approach we present here approach is directly extensible
to multidimensional free energy surfaces.  However, the numerical
details of performing the fitting may be challenging in some
cases. Both the optimization processes and the MCMC require successive
quadrature of the integrals $\int P_T(\vec{\xi} | \vec{\theta})
d\vec{\xi}$, which in all but the simplest cases cannot be carried out
analytically. The authors of vFEP have already noted this
challenge~\cite{Lee:JCTC:2014} in even two dimensions with splines. The mathematical approach presented 
in this paper may also be extensible to other methods that construct
biasing functions and FES adaptively, though the equations presented above will 
require modification if the sampling is not strictly stationary.

\clearpage
\section*{Appendix}

\subsection{Least squares functional fitting\label{sec:least_squares}}

One possibility briefly mentioned in the main text is to minimize a least squares fit of our trial function to the empirical distribution by writing the function to be minimized as
\begin{eqnarray*}
S(\vec{\theta}) &=& \int \left(P_E(\vec{\xi}|\{\vec{x}_n\}) - e^{-F(\vec{\xi}|\vec{\theta})}\right)^2 d\vec{\xi} \\
          &=& \int P_E(\vec{\xi}|\{\vec{x}_n\})^2 - 2P_E(\vec{\xi}|\{\vec{x}_n\}) e^{-F(\vec{\xi}|\vec{\theta})} \\
          && + e^{-2F(\vec{\xi}|\vec{\theta})} d\vec{\xi} \\
          &=& -2 \sum_{i=1}^N W(\vec{x}_n) e^{-F(\vec{\xi}_n|\vec{\theta})} \\
          & & + \int e^{-2F(\vec{\xi}|\vec{\theta})} d\vec{\xi}
\end{eqnarray*}
where we neglect the terms independent of $\vec{\theta}$ and employ eq.~\ref{eqn:expect} to estimate the thermal average.  However, this
integral is problematic as it is strongly biased towards low free energy regions. Large values of $F$ contribute very little to the sum or the log and are therefore largely unconstrained.

One could consider ameliorating this issue by minimizing over the relative error instead of the absolute. Since we can't divide by delta
functions, we would have to divide by the trial function:
\begin{eqnarray*}
S(\vec{\theta}) &=& \int \left(\frac{P_E(\vec{\xi}|\{\vec{x}_n\}) - e^{-F(\vec{\xi}|\vec{\theta})}}{e^{-F(\vec{\xi}|\vec{\theta})}}\right)^2 d\vec{\xi} \\
          &=& \int \left(P_E(\vec{\xi}|\{\vec{x}_n\})^2 e^{2F(\vec{\xi}|\vec{\theta})} \right. \\ 
          && \left. - 2P_E(\vec{\xi}|\{\vec{x}_n\}) e^{F(\vec{\xi}|\vec{\theta})} + 1\right) d\vec{\xi} 
\end{eqnarray*}
This integral is, however, even more problematic since squares of integrals of delta functions are not well-defined and the integral over the square of a delta function is infinite. In the
direct least squares approach, we didn't really care, because this
undefined function was independent of $\vec{\theta}$ and could be dropped, but in this case we must maintain this term. This seems an insurmountable deficiency and so we choose to abandon this approach.

Finally, we could consider minimizing over the squared log probabilities  (i.e the FES), instead of the weights.  This is \textit{not} the Kullback-Leibler divergence, but
does penalize divergence in the positive as well as the negative
direction:
\begin{eqnarray*}
S(\vec{\theta}) &=& \int P_E(\vec{\xi}|\{\vec{x}_n\}) \left(\ln \left(\frac{P_E(\vec{\xi}|\{\vec{x}_n\})}{P_T(\vec{\xi}|\vec{\theta})}\right)\right)^2 d\vec{\xi}\\
               &=& \int P_E(\vec{\xi}|\{\vec{x}_n\}) \left(\ln P_E(\vec{\xi}|\{\vec{x}_n\}) - \ln P_T(\vec{\xi}|\vec{\theta})\right)^2 d\vec{\xi}\\ 
               &=& \int P_E(\vec{\xi}|\{\vec{x}_n\}) \left(\ln P_E(\vec{\xi}|\{\vec{x}_n\})^2 \right. \\ 
               & & \left. - 2\ln P_E(\vec{\xi}|\{\vec{x}_n\}) \ln P_T(\vec{\xi}|\vec{\theta}) +\ln P_T(\vec{\xi}|\vec{\theta})^2\right) d\vec{\xi}
\end{eqnarray*} 
It appears that square minimizing the log weights isn't really
possible, because the logarithm of the empirical distribution of delta functions that
occurs in the cross-term is not well defined.  However, other least
square alternatives to determining similarities of distributions
involving the \textit{cumulative distribution} have been previously presented by
Schofield~\cite{Schofield:JPCB:2017}.

\subsection{Using biasing functions in conjunction with other accelerated sampling methods\label{sec:other_biasing}}

We can remove the requirement that the biasing functions are functions
of the collective variable, and simply assume that they are carried
out with different reduced potentials.  For the unbiased state
Kullback-Leibler divergence, eq.~\ref{eq:kldiverge} applies equally
well to any sampling, regardless of whether the additional samples
come from biases as a function of collective variables or not.

With more general potentials, the sample-weighted sum of biased Kullback-Leibler divergences is still computed as:
\begin{eqnarray}
\sum_{k=1}^{K} N_k D_{\mathrm{KL}}(\vec{\theta}) &=& \sum_{k=1}^K N_k \left(\sum_{n=1}^N W_k(\vec{x}_n) F_k(\vec{\xi}_n|\vec{\theta})\right. \nonumber \\
      && + \left.\ln \int e^{-F_{k}(\vec{\xi}'|\vec{\theta})} d\vec{\xi}'\right) \nonumber
\end{eqnarray}

To simplify this further, we first need to clarify what $\int
e^{-F_{k}(\vec{\xi}'|\vec{\theta})} d\vec{\xi}'$ means if the biasing
function is \textit{not} a function of $\vec{\xi}$.  In this case,
then there appears to be no clear relationship between
$F_k(\vec{\xi}|\vec{\theta})$ and $F(\vec{\xi}|\vec{\theta})$, so
information about $F_k(\vec{\xi}|\vec{\theta})$ will not help find a
best fit for $F(\vec{\xi}|\vec{\theta})$. So in the most general case,
one could only fit to a single unweighted empirical free energy surface of the
unbiased state, as shown in eq.~\ref{eq:kldiverge}.

However, there are circumstances when one could improve the overall
accuracy of the FES by performing a partial sum over only those of the
biased simulations that have umbrella sampling form, i.e. simulations
that have energy function of the form of eq.~\ref{eq:biased_u}, a sum
of the original $u(\vec{x})$ of interest and a bias function that only
depends on $\vec{\xi}$. Each of these umbrella sampling simulations
\textit{can} have many (say, $M_k$) simulations accelerated with other
methods associated with it, and we can use this information to build
our empirical estimate of the probability distributions of the $K$
biasing potentials. There are two primary situations we can consider.

First, reweighting is performed only between simulations that are
similar to the same umbrella sample, and they are reweighted to only
that particular one of the $K$ umbrella sampling simulations and no other
modifications. Each additional biasing simulation corresponds to exactly 
of the umbrella sampling simulations. 

Then these $K$ reweighted likelihoods are summed with some
$K$-dependent weights. In this case, there are $K$ different sets of
weights $W_k^{k'}(\x_n)$, one for each of the $K$ MBAR evaluations for
reweighting, where the subscripts denote that the weight is determined
for the $k$ simulations with biases alone, and the superscripts label
which set of weights they are.  For this situation, $N_k$ corresponds
to the total number of samples from all $M_k$ simulations associated
with the $K$th umbrella sampling potential.

We don't know what the optimal weights are for the $K$ reweighted
umbrella sampling likelihoods.  Because the number of effective number
of samples at any of the $K$ biased states will be less than $N_k$, we
replace the weighting $N_k$ with a constant $C_k$ to be determined
later. We then find:
\begin{eqnarray}
\sum_{k=1}^{K} C_k D_{\mathrm{KL}}(\vec{\theta}) &=& \sum_{k=1}^K C_k \left(\sum_{n=1}^{N_k} W_{k}^{k'}(\vec{x}_n) F_k(\vec{\xi}_n|\vec{\theta}) \right.\nonumber \\
 &+& \left. \ln \int e^{-F_{k}(\vec{\xi}'|\vec{\theta})} d\vec{\xi}'\right) \nonumber \\
                              &=& \sum_{k=1}^{K} \left(C_k \sum_{n=1}^{N_k} W_k^{k'}(\vec{x}_n) F(\vec{\xi}_n|\vec{\theta})\right) \nonumber \\
                              & & + \sum_{k=1}^{K} C_k \ln \int e^{-F(\vec{\xi}'|\vec{\theta})-b_k(\vec{\xi}')} d\vec{\xi}' \label{eq:sum_diff}\nonumber 
\end{eqnarray}
Where we have removed terms that are independent of the
parameters. Unlike for the derivation of eq.~\ref{eq:sumkldiverge}, we
cannot interchange the order of summation, and so there are no obvious
choices for $C_k$. One could choose an ``effective'' number of samples
that all of the samples from the $M_k$ simulation contribute to the $k$th umbrella sampling simulation.
for $C_k$, such as $\left[\sum_{n=1}^{N_k}
  W_k^{k'}(\vec{x}_n)\right]^{-2}$~\cite{Klimovich:JCAMD:2015}, though
it is not clear if this is optimal. However, eq. \ref{eq:sum_diff} is
still a usable equation to minimize divergence or as a log-likelihood.

In the second case, we assume that all $M = \sum_{k=1}^K M_k$
simulations are used to calculate a single set of MBAR weights
$W_k(\vec{x}_n)$ for each biasing function.  The additional biased
simulations are reweighted to all of the $K$ umbrella
sampling simulations. However, the normalization is a bit different
than is used in eq.~\ref{eq:sumkldiverge}. Although there is a single
$W_k(\vec{x}_n)$ corresponding to the weights in the $k$ biased
potentials, we cannot use the normalization $\sum_{k=1}^k N_k
W_k(\vec{x}_n) = 1$ to simplify the expression.  The equivalent
weighted sum here would have to be over all of the $M=\sum_k M_k$
states, and we are summing over only the $K$ states corresponding to
the $K$ umbrella sampling simulations.  We again use a weighted linear
scaling $C_k$ because the ``best'' weighting is not clear:
\begin{eqnarray}
\sum_{k=1}^{K} C_k D_{\mathrm{KL}}(\vec{\theta}) &=& \sum_{k=1}^K C_k \left(\sum_{n=1}^N W_{k}(\vec{x}_n) F_k(\vec{\xi}_n|\vec{\theta})\right. \nonumber \\
 & & + \left. \ln \int e^{-F_{k}(\vec{\xi}'|\vec{\theta})} d\vec{\xi}'\right) \nonumber \\
                               &=& \sum_{n=1}^N \left(\sum_{k=1}^K C_k W_k(\vec{x}_n)\right) F(\vec{\xi}_n|\vec{\theta}) \nonumber \\
                              & & + \sum_{k=1}^{K} C_k \ln \int e^{-F(\vec{\xi}'|\vec{\theta})-b_k(\vec{\xi}')} d\vec{\xi}' \label{eq:sum_same}
\end{eqnarray}

Where we have again removed terms independent of the
parameters. Eq.~\ref{eq:sum_same} is again somewhat more complex than
eq.~\ref{eq:sumkldiverge}, but usable as log-likelihood or a
divergence to minimize. One can again choose an ``effective'' number
of samples in the $k$th biased state for $C_k$, such as $C_k =
\left[\sum_{n=1}^{N} W_k(\vec{x}_n)\right]^{-2}$, though again it is
not entirely clear if this is optimal in any well-defined way.

\subsection{Efficient minimization of splined surfaces\label{sec:derivatives}}

We briefly describe efficient optimization routines to solve the minimization problems defined in eqs.~\ref{eq:max1} and~\ref{eq:max2} in the case of splines. In below, we suppress explicit dependence of F on $\theta$ for compactness. We start by examining the minimization of eq.~\ref{eq:max2}:
\begin{equation*}
S(\theta) = \sum_{n=1}^N F(\vec{\xi}_n) + \sum_{k=1}^{K} N_k \ln \int e^{-F(\vec{\xi}')-b_k(\vec{\xi}')} d\vec{\xi}' - \ln \mathcal{P}(\vec{\theta})
\end{equation*}
Various minimization approaches are required to compute the gradient and Hessian of this function with
respect to the parameter vector $\vec{\theta}$. 
For convenience, we define the equilibrium average performed with biasing function $k$ of some observable $A$ that is a function of $\vec{\theta}$ as:
\[ \langle A(\vec{\theta}) \rangle_k = \frac{\int A(\vec{\xi}' | \vec{\theta}) e^{-F(\xi',\theta)-b_k(\vec{\xi}')} d\xi'}{\int e^{-F(\vec{\xi}',\theta)-b_k(\vec{\xi}')}d\vec{\xi}'}\]
The $i$ components of the gradient are then:
\begin{equation*}
\nabla S(\theta)_i = \sum_{n=1}^N\frac{\partial F(\vec{\xi})}{\partial \theta_i} \nonumber + \sum_{k=1}^K N_k \left\langle \frac{\partial F(\vec{\xi}')}{\partial \theta_i}\right\rangle_k - \frac{1}{\mathcal{P}(\theta)} \frac{\partial \mathcal{P}(\theta)}{\partial \theta_i}
\end{equation*}
We note that if we have linear basis functions, the first term is
independent of $\vec{\theta}$ and can be precomputed, as $\frac{\partial F}{\partial \theta_i}$ is simply the
corresponding basis function. Additionally, the integral term will
have only limited support for each basis function, so the integrals
are relatively easy to carry out, and the calculations scales easily
in the number of basis functions.

%
The $ij$ entries in the Hessian are::
\begin{eqnarray}
\nabla^2 S(\theta)_{ij}                             &=& \sum_{n=1}^N \frac{\partial^2 F(\vec{\xi}_n)}{\partial \theta_i \partial \theta_j} \nonumber \\
                              & & - \sum_{k=1}^{K} N_k \left[\left\langle \frac{\partial^2 F(\vec{\xi})}{\partial \theta_i\partial \theta_j}\right\rangle_k - \left\langle\frac{\partial F(\vec{\xi})}{\partial \theta_i} \frac{\partial F(\vec{\xi})}{\partial \theta_j}\right\rangle_k \right.\nonumber \\
                              & & \left.+ \left\langle\frac{\partial F(\vec{\xi})}{\partial \theta_i} \right\rangle_k \left\langle\frac{\partial F(\vec{\xi})}{\partial \theta_j} \right\rangle_k\right] \nonumber \\
& - &  \left[\frac{1}{\mathcal{P}(\theta)} \frac{\partial \mathcal{P}(\theta)}{\partial \theta_i\partial \theta_j} - \frac{1}{\mathcal{P}(\theta)^2} \frac{\partial \mathcal{P}(\theta)}{\partial \theta_i}\frac{\partial \mathcal{P}(\theta)}{\partial \theta_j}\right]
\end{eqnarray}
If we assume that we have a trial function that is linear in the
parameters, then the initial terms involving mixed second derivatives vanish, leaving only:
\begin{eqnarray}
\nabla^2 S(\theta)_{ij} &=& \sum_{k=1}^{K} N_k \left[\left\langle \frac{\partial F(\vec{\xi})}{\partial \theta_i} \frac{\partial F(\vec{\xi})}{\partial \theta_j}\right\rangle_k \right.\nonumber \\
                   & & \left. - \left\langle\frac{\partial F(\vec{\xi})}{\partial \theta_i} \right\rangle_k \left\langle\frac{\partial F(\vec{\xi})}{\partial \theta_j} \right\rangle_k\right] \nonumber \\
& - &  \left[\frac{1}{\mathcal{P}(\theta)} \frac{\partial \mathcal{P}(\theta)}{\partial \theta_i\partial \theta_j} - \frac{1}{\mathcal{P}(\theta)^2} \frac{\partial \mathcal{P}(\theta)}{\partial \theta_i}\frac{\partial \mathcal{P}(\theta)}{\partial \theta_j}\right]
\end{eqnarray}

If the function is linear in the parameters (again, such as splines), this will only be nonzero in areas where basis functions have mutual support, essentially just banded along the diagonal, so are be relatively inexpensive to compute.

In the case of eq.~\ref{eq:max1}, this becomes: 
\begin{eqnarray}
\nabla S(\theta)_i = N \sum_{n=1}^N W_n(\vec{x}_n) \frac{\partial F(\vec{\xi})}{\partial \theta_i} - N \left\langle \frac{\partial F(\vec{\xi}')}{\partial \theta_i}\right\rangle - \frac{1}{\mathcal{P}(\theta)} \frac{\partial \mathcal{P}(\theta)}{\partial \theta_i} \nonumber \\
\end{eqnarray}
\begin{eqnarray}
\nabla^2 S(\theta)_{ij} &=& N\left(\left\langle \frac{\partial F(\vec{\xi})}{\partial \theta_i} \frac{\partial F(\vec{\xi})}{\partial \theta_j}\right\rangle\right. \nonumber \\
& & - \left.\left\langle\frac{\partial F(\vec{\xi})}{\partial \theta_i} \right\rangle \left\langle\frac{\partial F(\vec{\xi})}{\partial \theta_j} \right\rangle \right)\nonumber \\
& - &  \left[\frac{1}{\mathcal{P}(\theta)} \frac{\partial \mathcal{P}(\theta)}{\partial \theta_i\partial \theta_j} - \frac{1}{\mathcal{P}(\theta)^2} \frac{\partial \mathcal{P}(\theta)}{\partial \theta_i}\frac{\partial \mathcal{P}(\theta)}{\partial \theta_j}\right] \nonumber \\
\end{eqnarray}
Where expectations are now over the \textit{unbiased} state rather than any of the $K$ biased simulations.

\bibliography{zotero}
\newpage
\end{document}